\documentclass{statsoc}
\usepackage[a4paper]{geometry}
\usepackage{graphicx}
\usepackage{caption}
\usepackage{subcaption}
\expandafter\def\csname ver@subfig.sty\endcsname{}
\usepackage[textwidth=8em,textsize=small]{todonotes}
\usepackage{amsmath}
\usepackage{natbib}
\usepackage{subfig}
\usepackage{amssymb}
\usepackage{bbold}
\usepackage{breqn}
\usepackage{mleftright}
\usepackage{diagbox, eqparbox, hhline}
\usepackage{xcolor, soul}
\soulregister\cite7
\soulregister\ref7
\soulregister\pageref7

\newcommand{\Pro}{\mathbb{P}}
\newcommand{\N}{\mathbb{N}}

\usepackage{hyperref}
\hypersetup{colorlinks,allcolors=black}

\title[Test allocation based on risk of infection]{Test allocation based on risk of infection from first and second order contact tracing}
\author[Gabriela Bayolo Soler {\it et al.}]{Gabriela Bayolo Soler, Miraine Dávila Felipe, Ghislaine Gayraud}
\address{Université de technologie de Compiègne, LMAC (Laboratory of Applied Mathematics of Compiègne), CS 60 319 - 60 203 Compiègne Cedex} 
\email{gabriela.bayolo-soler@utc.fr}

\begin{document}

\begin{abstract}

Under limited available resources, strategies for mitigating the propagation of an epidemic such as random testing and contact tracing become  
inefficient. Here, we propose to accurately allocate the resources by computing over time an individual risk of infection based on the partial observation of the epidemic spreading on a contact network; this risk is defined as the probability of getting infected from any possible transmission chain up to length two, originating from recently detected individuals. 
To evaluate the performance of our method and the effects of some key parameters, we perform comparative simulated experiments using data generated by an agent-based model. 

\medskip

\end{abstract}

\section{Introduction}

In the context of epidemics, contact tracing is the process of identifying individuals who have been in contact with other individuals diagnosed with a transmissible disease. The relevant contacts are those that would allow the transmission to happen, which depends on the mode of transmission of the disease, and requires the detected individual to be infectious at the time of the encounter.
Together with strategies such as testing and quarantining, contact tracing has been shown to be an essential mechanism in order to mitigate the spread of a disease, allowing to contain and delay outbreaks, see \cite{brandt2022history}.
Ideally, detected individuals (those who receive a positive test result) are quarantined and their contacts are identified and tested as well. 
However, these interventions have an economic and social cost, and a scenario where all the contacts of detected individuals are tested is not realistic for diseases that start spreading quickly in the population (outbreak). Hence, when the resources are limited (e.g. amount of daily available tests), the question of how to cleverly allocate them to the population arises. 

In this direction, we propose a method to compute the risk of infection of individuals in the population over time, based on the partial observation of the epidemic spreading through the population contact network. The risk of each individual is defined as her/his (marginal) probability of infection conditionally on the observed variables in the recent past, and the higher-risk individuals can get notified to be tested, quarantined, or applied any other preventive measures. Thus, the quantification of the infection risk is proposed here as a tool to allocate the available resources more rationally than just randomly. Similar intervention approaches have been shown to have a positive impact on the mitigation of epidemics, see for instance \cite{baker2021epidemic, herbrich2020crisp, Batlle22, 2023notimetowaste, guttal2020risk, viratrace, murphy2021risk, gupta2023proactive, bengio2020predicting, sattler2020risk, alsdurf2020covi}. 

To be more precise about our approach, we consider an agent-based model on a fixed-size population where individuals admit a set of (discrete) characteristics influencing the transmission of the disease. The pairwise contacts between individuals are described by a \textit{dynamical network} model, in which some connections are deleted and some others are created while time evolves. 
Then, we depict the disease spread in the population contact network by a stochastic \textit{Susceptible-Infectious-Removed} (SIR) dynamic (including more than three classes), meaning that contagion can only happen (with a certain probability) when an infectious individual is connected with a susceptible one by an edge in this contact network.
In particular, we consider a non-Markovian dynamic since the infection probability per interaction depends on the date of infection of the source. This probability is also a function of the previously mentioned characteristics of both individuals and the context of the interaction (e.g. in the household, workplace, or random).

Given this dynamical network model and the propagation process on this network, we suppose that the infectious statuses of individuals are (partially) observed through testing, as well as the underlying contact network, and the factors having an impact on the transmission for the set of tested individuals and their direct and secondary contacts.
Actually, we consider \textit{at risk} not only the first-degree contacts of detected individuals ($1^\circ$ contacts) but also their subsequent contacts ($2^\circ$ contacts).
In the sequel, to distinguish the different groups of individuals under consideration, we  call \textit{index cases} the  detected individuals who are infectious, \textit{$1^\circ$ contacts} the individuals who are not detected but interacted with index cases while the latter were infectious,  and, \textit{$2^\circ$ contacts} the individuals who are not detected but interacted with $1^\circ$ contacts after the latter were in contact with index cases. In addition,  \textit{$1^\circ$ interaction} and \textit{$2^\circ$ interaction} refer to the risky encounter between index cases and $1^\circ$ contacts, and between $1^\circ$ and $2^\circ$ contacts respectively.

In real situations, the information about the tested individuals and their $1^\circ$ and $2^\circ$ contacts is provided either by individuals themselves, through manual contact tracing (MCT), or by a digital contact tracing (DCT), and is not necessarily centralised in the digital case. The advantages of extending the contact tracing strategies up to $2^\circ$ contacts have been highlighted in the recent literature, see \cite{viratrace,Firth20,Weigl21}.
Then, to compute the risk, we consider a rather general probability of transmission per interaction, depending on the observed attributes of both individuals, characteristics of the interaction, and the infection time of the source (which is unknown and estimated from the observations). 
Finally, we compute the probability for each individual at risk of having been infected by one of the individuals detected in the previous days (fixed time window), through chains of transmission of length one or two. It is worth noticing that we compute this marginal probability by summing over all the possible paths of transmission, providing explicit formulas, and avoiding independence assumptions and the \textit{cycling back} phenomenon described in Section 2 of the Supplementary Material of this paper (see \cite{supplementary_material}).

Finally, we propose and simulate the following mitigation strategy: every day the probability of being infected is computed for the individuals at risk, and a fixed number of the highest-ranked individuals are tested; the newly detected individuals are put in quarantine the day after, and the process is repeated each day during an intervention period.
In parallel of the detection by risk, symptomatic individuals are tested with a fixed probability per day since the beginning of their symptoms, the ones detected are quarantined, and their contacts are traced as described before.
We evaluate this intervention through simulations in a series of different scenarios, where we study the impact of some of the parameters in the model such as the number of daily available tests, the time-frame for the time of infection and the proportion of detection of symptomatic individuals. We also study the influence of the probability function of transmission per interaction, and different ways to estimate the time of infection of the source. We found that, in most cases, our test allocation method is capable to  mitigate the propagation of the disease considerably faster than randomly selecting (RS) individuals to get tested, or the usual contact tracing (CT, i.e. ranking according to the number of interactions with detected individuals). 
Moreover, we found that with fewer daily available tests, our risk ranking is more efficient than an equivalent setting where the probabilities are computed under the mean-field (MF) hypothesis, see \cite{baker2021epidemic}.

In our simulations, we use the agent-based model Oxford OpenABM-Covid19, which defines a contact network based on demographic characteristics of the UK population calibrated for the transmission of airborne diseases, \cite{hinch2021openabm}. Moreover, in this model, the spread of the COVID-19 epidemic follows an enriched SIR dynamic (11 possible disease statuses), stratified by age group and context of the interaction (household, workplace, random encounter), and considering the main epidemiological characteristics of COVID-19 outbreak for the parameter calibration. In particular, the model defines an infectiousness function that depends on the time since the infection and takes into account asymptomatic/pre-symptomatic disease states. In conclusion, this model captures essential features of the contacts in real populations, as well as the real epidemiology of the COVID-19 pandemic.

In the recent literature, there are many research works that study the effect of contact tracing combined with treatment and/or quarantine, as non-pharmaceutical interventions for infectious disease mitigation and control, see \cite{eames2003contact}. The SARS-CoV-2 outbreak has considerably increased the scientific research motivation around strategies including contact tracing. In particular, DCT apps have attracted the attention of Public Health authorities and the scientific community as well, \cite{Cebrian21}. DCT apps allow to collect the information automatically, and provide fast processing times but also come with drawbacks regarding privacy and the data protection standards ruling in most countries, see \cite{zwitter2020big}. It is not our aim to discuss here if the contact tracing strategies should be implemented manually or through digital mobile apps, and we suppose in what follows that the tracing of contacts is done within one day after detection. However, it is worth noticing that the tracing of random contacts, included in our simulations, is only possible through DCT.
The effectiveness of contact tracing interventions for the COVID-19 pandemic is studied in  \cite{jenniskens2021effectiveness} and more recently in \cite{Pozo23}, providing reviews of what has been done on this topic based on empirical and simulated data. 

Among the recent research studies looking at non-pharmaceutical interventions strategies, there are a few with the same aim as ours, that is, to integrate individual infection risk levels based on the observation of the interaction network and the test results in order to optimise the allocation of the available resources. 
In \cite{viratrace} and \cite{Batlle22}, the risk is computed using Monte Carlo methods. In \cite{viratrace}, the authors estimate the individual infection probability up to $3^\circ$ contact tracing, arguing that it improves the detection of asymptomatic patients in diseases with a high percentage of them. Here, instead, we derive an explicit formula for these probabilities taking into account up to $2^\circ$ contacts, and hence avoiding the large computing power and the centralised information required by Monte Carlo methods. Other works such as \cite{baker2021epidemic} and \cite{guttal2020risk} avoid the use of Monte Carlo methods by using the mean-field approximation to evaluate the individual risk of infection. However,  the way the risk is propagated is ``bidirectional'', meaning that it is not only ``forward'' in the direction of the transmission  given the observations. Indeed, they suppose that individuals interchange their risk information at each time step if there is an edge between them, regardless of the previous path followed by the transmitted risk, falling sometimes in the above mentioned cycling back issue.
Despite the similarities in the use of the MF hypothesis, it should be noticed that in \cite{guttal2020risk} the network and propagation model are simpler than in \cite{baker2021epidemic}. The latter deals with more realistic models, including the OpenABM-Covid19 model used in our simulations. 
Moreover, in \cite{baker2021epidemic} a second method is developed, that estimates the individual infection risk as the posterior distribution conditional on the test observations through the Belief Propagation (BP) inference algorithms. Similar computations are achieved in \cite{herbrich2020crisp} and \cite{2023notimetowaste} using Gibbs Sampling (GS) and Factorized Neighbors (FN) respectively.
Another related methodology is presented in a series of works \cite{bengio2020predicting,gupta2023proactive,alsdurf2020covi, biazzo2022bayesian, shah2020finding, deepdemixing,tomy2022estimating, tan2023deeptrace}, where the authors achieve the risk computation using deep learning algorithms based on neural networks. It is also worth mentioning other machine learning approaches for risk estimation from DCT data that exploit different observations, such as Bluetooth energy measurements and exposure data, see \cite{sattler2020risk,murphy2021risk}.

A crucial aspect of our work is that we consider a rather realistic contact network and a detailed disease spread model, see \cite{hinch2021openabm}. Another significant aspect is that we consider $2^\circ$  contact tracing, providing a more accurate estimation of the risk compared with the $1^\circ$ contact tracing. Although our approach can be extended to $3^\circ$ contacts and beyond, the calculations would get much heavier, and we argue that the gain in the effectiveness of the mitigation would not be significant, due to the uncertainty on the statuses of the intermediate individuals in the chains of transmission. One more core feature of our method is that to compute the risk, neither the whole contact network nor a centralised setup (contacts and individual information) is required.  These characteristics are essential for the practical implementation using DCT applications, where the interchange of information between contacts over the whole network could be difficult due to a huge amount of personal data impacted by privacy restrictions. These difficulties can be intensified in the centralised case, see \cite{2023notimetowaste}. Furthermore, compared to the previously mentioned inference algorithms used to calculate the individual risk of being infected (i.e. BP, GS, FN), our risk calculation is simpler:
while these algorithms integrate the observations at any time $t$ by updating and re-propagating the risks step-by-step in a given time interval previous to $t$ for every contact (up to any contact degree) of all the individuals in the population, we calculate directly the risk at $t$ of individuals in contact (up to $2^\circ$) with someone detected by integrating the probability of any possible path of length up to 2 that might lead to the infection of these individuals. In this way, we avoid any cycling back phenomenon, and we do not need to update the risk of all individuals for every time step in the contact tracing time window, getting a very low level of messages interchange between individuals, see \cite{2023notimetowaste}.

We define our approach in Section \ref{sec:methods}. In Section \ref{sec:simulation_results} we study, compare and discuss the impact of our intervention method when applied to simulated data for COVID-19 epidemic, using the OpenABM-Covid19 model. We finish with a discussion in Section \ref{discussion}. 
In the Supplementary Material we provide more details on the agent-based model used in our simulations, and we compare, through a toy example, the MF approximation with our approach.

\section{Methods}\label{sec:methods}

We aim at defining a practical, realistic, and efficiently  implemented  risk-based dynamic detection process allowing us to identify the most likely infected individuals. 
To take into account the heterogeneity of disease transmissions,  we consider that the probability of infection per interaction depends on individual attributes (e.g. age, healthy habits) and infectiousness of the source (e.g. day since infection, type of symptoms)  of individuals in contact and the characteristics of the interaction (e.g. place, duration, distance, protective measure). 

In addition, to provide a sharper and earlier detection process of the most likely infected individuals, not only the direct contacts of detected individuals are considered to be at risk, but also their subsequent contacts ($2^\circ$ contact tracing). 
Compared with the usual $1^\circ$ contact tracing, we expect to obtain a more accurate estimation of the risk of infection, and hence to detect more efficiently (in terms of the mitigation of the epidemic) individuals that are in general harder to detect due to the absence of symptoms (asymptomatic or pre-symptomatic individuals). It has become clear from numerous research studies that these latter individuals play an important role in SARS-CoV-2 transmission dynamics, see \cite{ferretti2020quantifying,Weigl21,Firth20}.

From a realistic point of view, one can estimate the individual risk of getting infected only from the observations (available information) that are provided by the tested individuals and their contacts through manual or digital CT.
In view of all the above, our intervention approach relies on a dynamic risk evaluation for $1^\circ$ and $2^\circ$ contacts, and their risk is defined as the marginal conditional probability of getting infected given their past known $1^\circ$ and $2^\circ$ interactions, and the information provided by the tests.

In the following, we give a brief description of disease-spread models on social networks,  we introduce some useful notations and finally, we define the risks of infection  for $1^\circ$ and $2^\circ$ contacts.

\subsection{Disease spread model on social networks}

We consider a population consisting of $N$ ($N\in\mathbb{N}$) individuals that stays constant over time, so neither births, deaths nor  migrations are taken into account. Notice that constant population size is a convenient assumption for the notation, but small variations in the population size do not influence the procedures described in the sequel. 
Let us denote by $V=\{1,\ldots,i,\ldots,j,\ldots,N\}$ the population under consideration.

\bigskip
\textit{Social structure model}

At any discrete time $t$ ($t\in \mathbb{N}$), the social structure (interactions between individuals at $t$) is represented by an undirected graph $\mathcal{G}_t  = \left( \mathcal{V}, \mathcal{E}_t  \right) $. We consider that $\mathcal{V} = \{ (i, a^i) : i \in V \text{ and } a^i \in A \}$  corresponding to  the set of vertices  $V$ (the individuals), supplemented by the set $A$ of vertices attributes mentioned at the beginning of the section. Likewise, $\mathcal{E}_t = \{ (i,j,c^{ij}_t) : (i,j) \in E_t \text{ and } c^{ij}_t \in C \}$ is the set of the edges $E_t$,  describing the interactions at time $t$ between the corresponding individuals, supplemented by $C$, the set of the characteristics of these interactions. 

The time interval $[0:T]$ corresponds to the period of study, where we assume that at the first time $0$ there is already an ongoing outbreak (a small number of infectious individuals) and at the last time $T$ the study ends. Here the time unit is one day.  In the sequel, to refer to the discrete-time interval between any $t_l$ and $t_m$, we use either $\left[ t_l:t_m\right]$ with the convention that the interval is empty when $t_l>t_m$. 

Although the stochastic mechanism of the social network evolution over time is not of primary interest here,  the sequence $\mathcal{G}_{[0 : T]} $  is however viewed  as a realisation of a dynamic random network model over the time-period $[0: T]$ (see \cite{britton2020epidemic}). As previously mentioned, it is reasonable to consider that the social structure is partially random (except the sub-graph corresponding to household interactions), and hence that some individual and edge attributes are governed by some specific probability distributions.

\bigskip
\textit{Infectious disease spreading on the network}

Here, we consider an individual-based SIR dynamic spreading on the underlying social network. The possible individual statuses are only Susceptible ($S$), Infected ($I$) and Removed ($R$), and the only possible status evolution over time are $S \rightarrow I$ and $I \rightarrow R$, where $R$ considered as an absorbing state. We denote by $X^i_t\in \{ S, I, R \}$ the random variable corresponding to the status of individual $i$ at time $t$.

While the dynamic social network is represented by undirected graphs, the transmission of the infectious disease is directed along an edge from an infectious individual (\textit{source} or \textit{donor}) to a susceptible one (\textit{recipient}).
We consider that the transmission probability depends on the infection time of the source, as well as on both  individual and interaction attributes $A$ and $C$.
For any time $t\geq 0$, and any two distinct individuals $i$ and $j$ in $V$, seen respectively as possible source and recipient, let us denote by  $a^{i}, a^{j} \in A$  the individual attributes, and  by $c_t^{ij} \in C$ the characteristics of the interaction. In addition, we consider another attribute related to the disease spreading, namely $b_t^i \in B$, which corresponds to  the type of symptoms the individual $i$ manifests at time $t$. We have in mind that the individual $i$ could be for example asymptomatic, mild or severe, and the parameter $b_t^i$ allows us to establish how the severity of the symptoms influences the probability of transmission of $i$. Finally, we denote by $\tau_I^i$ and $\tau_R^i$  the not observed random variables representing the infection and removal times of individual $i$, taking values in $[0:t] \cup \{ +\infty \}$, where  
for convenience we set $\tau_I^i =  \tau_R^i = \infty$ if  $i$ has not yet been infected.
We suppose that the transmission probability that the individual $i$ infects $j$ at $t$ depends on all the above quantities (parameters and random variables). Hence, we denote it by $\lambda^{i\rightarrow j}_{a^{i}, a^{j}, t, b_t^{i},   \tau_I^i, \tau_R^i,c_t^{ij} }$ and for the  sake of simplicity we use the following  simplified notation,

 \begin{align*}
  \lambda^{i\rightarrow j}_{a^{i}, a^{j}, t, b_t^{i},   \tau_I^i, \tau_R^i,c_t^{ij} }  \equiv \lambda^{i\rightarrow j}_{\tau_I^i, t}. \label{Proba_Trans}
  \end{align*}

Notice that 
\begin{equation}
    \lambda^{i\rightarrow j}_{\tau^i_{I},t}
 = 0  \quad \text{if}   \quad \left(i,j \right) \notin E_t \quad \text{or} \quad t\leq  \tau^i_{I} \quad \text{or} \quad t\geq  \tau^i_{R}.
\label{e:cond_prob}
\end{equation}
 
\textit{Observations}

Let us now describe the observation process on which the risk computation relies. 
We assume that testing individuals for the disease is possible from time $t=1$, so for any $t \ge 1$ let us denote by $\mathcal{D}^+_t $,  $\mathcal{D}^-_t $ the set of individuals receiving a positive, respectively negative result at time $t$.
In order to focus on the most recent and relevant interactions, we introduce the parameter $\zeta\in \mathbb{N}$ corresponding to the contact tracing time-frame. Hence, at a given time $t\ge 0$, the set of observations is provided by the graph of interactions during the recent days $[t- \zeta: t]$, and the set  of individuals with a positive and a negative result until $t-1$. This set includes test results, interactions network, and both individual and interaction attributes.
Notice that we keep the list of all the individuals detected since the beginning because we consider that after they get the infection, they stay immune for the period of study.
In conclusion, the set of observations at $t$ is defined as,

\begin{equation*}
    \mathcal{O}^{\zeta}_t = \left\lbrace \mathcal{G}_l\right\rbrace_{ l \in [t-\zeta : t]} \cup \left\lbrace \mathcal{D} ^+_l \right\rbrace _{ l \in [1 : t-1]}  \cup \left\lbrace \mathcal{D} ^-_l \right\rbrace_{ l \in [1 : t-1]}.
\end{equation*} 

In practice, the spreading of the disease on the network is not available. Unless tested or with reported symptoms, the status and infectiousness of individuals are unknown. In addition, there is some uncertainty in the detection process due to the sensitivity and specificity of the tests, and the co-circulation of other diseases causing similar symptoms. However, here we consider only perfect tests (i.e. test specificity and sensitivity equal to 1) and we assume that all the reported symptoms are a consequence of the disease under study. 

In particular, the infection and removal times of individuals are never known even for the detected individuals. 
These quantities are necessary to compute the probability of transmission from a possibly infectious individual $i$ to a presumably susceptible $j$ at time $t$, as well as the severity coefficient $b_t^i$, therefore we approximate them.
More precisely, we approximate the probability distributions of $\tau^i_I$ and $\tau_R^i$, and for the sake of simplicity, we keep the notations $\tau^i_I$  and $\tau^i_R$ to represent the random variables issued from the approximated distributions.
Depending on the proposed contact tracing approach, we approximate the distribution of $\tau_I^i$ by a Dirac measure $\delta_{\widehat{\tau}^i_{I}}$ or a generalised truncated geometric distribution, as seen later in Equations \eqref{e:law_dirac-infect} and \eqref{e:law-geo-infect}. Similarly, 
we approximate the distribution of $\tau_R^i$ by a Dirac distribution $\delta_{\widehat{\tau}^i_{R}}$.
For convenience, the quantities $\widehat{\tau}^i_{I}$  and $\widehat{\tau}^i_{R}$ are defined in $\N \cup \{+\infty\}$ and serve as  estimations  of $\tau_I^i$, $\tau_R^i$. 
By default, at time 0, we set $\widehat{\tau}^i_{I}=\widehat{\tau}^i_{R}=+ \infty$ for all $i\in V$. We briefly describe two distinct situations that we have at any time $t$.
 
\begin{enumerate}
    \item If $i\in \mathcal{D}^+_t$, we update the values of the estimators $\widehat{\tau}^i_{I}$ and $\widehat{\tau}^i_{R}$ to finite values computed from the observations. The details on the definition of the finite candidates for $\widehat{\tau}^i_{I}$ are provided in Section \ref{ssec:estimation_time_infection}. We set $\widehat{\tau}^i_{R}=\widehat{\tau}^i_{I}+L$, where $L$ is a positive integer that is chosen greater than the mean duration of infectiousness ($L=21$ in our simulations). 
    \item If $i \notin \mathcal{D}^+_t$, the values of $\widehat{\tau}^i_{I}$ and $\widehat{\tau}^i_{R}$ stay equal to their previous values.  
\end{enumerate}

We denote by $\widehat{b}_t^i$ the estimation of the parameter $b_t^i$ at time. If $i\notin \mathcal{D}_{[1:t]}^+$, we consider $\widehat{b}_s^i= \text{``asymptomatic''}$ for $0 \le s\le t$. On the other hand, if $i$ is detected at time $t$, that is $i\in \mathcal{D}^+_t$, we update $\widehat{b}_l^i$ as the real value $b_t^i$ for $l\in [ \widehat{\tau}^i_{I}, \widehat{\tau}^i_{R}]$ since we assume that when an individual is detected, the severity of the symptoms experienced by this individual is known.

To keep the trace of the negative test results, we define at any $t$ and for any individual $j$,  the day $\theta^j_{t}$ as the last date prior to $t$ on which  $j$ receives a negative test result,

\begin{equation*}\label{1-order-start-time}
\theta^j_{t} =  \max \left\lbrace l \in [1:t]   : j\in \mathcal{D}^{-} _{l} \right\rbrace,
\quad \text{and by convention }\; \max \emptyset = 0. 
\end{equation*}
 
Hence, only the interactions of $j$ that are posterior to $\theta^j_{t}$ are considered risky, meaning that, before a negative test result, the probability that $j$ has been infected is zero. 

\subsection{Risk of infection via transmission chains}\label{ssec:risk}

We propose two approaches to compute the risk, based on two different degrees of interactions. To differentiate both methods, we call them in the sequel $1^\circ$contact tracing ($1^\circ$CT) and $2^\circ$contact tracing ($2^\circ$CT). In the first approach, the risk of infection is based on $1^\circ$ interactions, while the second proposes a more accurate risk of infection, defined from both $1^\circ$ and $2^\circ$ interactions. 

For any time $t$ and any individual $j\notin {\mathcal D}^+_{[1:t-1]}$, our aim is to estimate the probability of infection of $j$  given the set of observations $\mathcal{O}^{\zeta}_t$, that is

\begin{equation}
    \Pro \left(X^j_t = I \; \middle| \;  \mathcal{O}^{\zeta}_t\right).
\label{e:aim_prob}
\end{equation}

We introduce a new truncation parameter $\gamma \in \N$, such that $\gamma \leq \zeta$, corresponding to the infection time-frame of interest. 
More precisely, for any $j \notin {\mathcal D}^+_{[1:t-1]}$, we approximate the probability in \eqref{e:aim_prob} by ``$ R^j_{\gamma, \zeta} \left( t \right)$'', which is defined as the probability of $j$ being infected in the interval $[t-\gamma : t]$ given the set of observations at time $t$, that is

\begin{equation}
    R^j_{\gamma, \zeta} \left( t \right)  = \Pro \left(\bigcup_{l = t - \gamma}^t \left\lbrace \tau^j _I = l \right\rbrace \; \middle| \;  \mathcal{O}^{\zeta}_t \right).
    \label{e:risk}
\end{equation}

The risk given by Equation \eqref{e:risk} can be expressed as the probability for individual $j$ of having been infected by any of the possible sources of transmission in the time interval  $[t-\gamma : t]$, given the observations. Hence, Equation \eqref{e:risk} can be rewritten as

\begin{equation}
    R^j_{\gamma, \zeta} \left( t \right) = 1 - \prod_{i \in V : i\neq j} \Pro \left( \bigcap_{l = t - \gamma}^t \left\lbrace Y^{ij}_l = 0\right\rbrace  \; \middle| \; \mathcal{O}^{\zeta}_t \right), 
    \label{e:risk_2}
\end{equation}

\begin{align*}
\text{where} \quad & Y^{ij}_l =
\begin{cases}
1, \text{ if $i$ infects $j$ at time $l$,}\\
0, \text{ otherwise. }
\end{cases}
\end{align*}

Then, for any possible individual $i$ considered as a source, we can use the law of total probability with respect to the date of infection of $i$, which leads to

\begin{equation}
     \Pro \left( \bigcap_{l = t - \gamma}^t \left\lbrace Y^{ij}_l = 0 \right\rbrace \; \middle| \;  \mathcal{O}^{\zeta}_t \right) = \sum_{d\in \{1,2,..., t-1, \infty \} } \Pro \left( \bigcap_{l = t - \gamma}^t \left\lbrace Y^{ij}_l = 0 \right\rbrace \; \middle| \;  \tau^i_I = d, \mathcal{O}^\zeta _t  \right)  \Pro \left(\tau^i_I = d \; \middle| \;  \mathcal{O}^{\zeta}_t \right).
     \label{e:total_prob}
\end{equation}

By independence of the events $\left\lbrace Y^{ij}_l = 0 \; \middle| \;  \tau^i_I = d, \mathcal{O}^{\zeta}_t \right\rbrace _{l\in[t-\gamma : t]} $ we have

\begin{equation*}
   \Pro \left( \bigcap_{l = t - \gamma}^t \left\lbrace Y^{ij}_l = 0 \right\rbrace \; \middle| \;  \tau^i_I = d, \mathcal{O}^{\zeta}_t \right) = \prod_{l = t - \gamma}^t \Pro \left(  Y^{ij}_l = 0 \; \middle| \;  \tau^i_I = d, \mathcal{O}^{\zeta}_t \right)  ,
\end{equation*}

where we assume,
\begin{equation*}
   \left. Y^{ij}_l \; \middle| \; \tau^i_I = d, \mathcal{O}^{\zeta}_t \right.  \; \sim \; \text{Ber} \left(\lambda ^{i\rightarrow j}_{d,l} \mathbb{1} \left( l > \theta ^j _{t} \right) \right).
\end{equation*}

The way we model $\left.\tau^i_I  \; \middle| \;  \mathcal{O}^{\zeta}_t \right.$  depends on the contact tracing method under consideration and it is explained in Section \ref{ssec:1-CT} and  Section \ref{ssec:2-CT}. 

\subsubsection{Infection risk for 1°contact tracing}\label{ssec:1-CT}

We describe now how to compute the risk defined by Equation \eqref{e:risk_2} for any $j$ such that $j \notin {\mathcal D}^+_{[1:t-1]}$ using the $1^\circ$CT approach that considers as possible sources of infection only the index cases.  
At a given time $t$, an individual $i$ that has been detected up to $t$, is considered as an \textit{index case} for any time $l$ between the respective estimated infection and removal times. Thus, we define the set of index cases at time $l$ given the set of observations $\mathcal{O}_t^{\zeta}$ as

\begin{equation*}
    \mathcal{I}_{l,t} = \left\lbrace i\in \mathcal{D}^+_{[1:t-1]} : l \in ]\widehat{\tau}^i_{I} : \widehat{\tau}^i_{R}[ \right\rbrace.
\end{equation*}

For the $1^\circ$CT approach, we only take into account the interactions with index cases that occur in   the interval $[t - \gamma: t]$. Consequently, the set of observations reduces to 

\begin{equation*}
    \mathcal{O}^{\gamma, 1^\circ }_t = \left\lbrace \mathcal{G}_{l,t}^{1^\circ}\right\rbrace_{ l \in [t-\gamma : t]} \cup \left\lbrace \mathcal{D} ^+_l \right\rbrace _{ l \in [1 : t-1]}  \cup \left\lbrace \mathcal{D} ^-_l \right\rbrace_{ l \in [1 : t-1]} \subset \mathcal{O}^{\zeta}_t,
\end{equation*} 
where $\mathcal{G}_{l, t}^{1^\circ} = \left( \mathcal{V}_{l, t}^{1^\circ} , \mathcal{E}^{1^\circ}_{l, t}  \right) \subset \mathcal{G}_l$,  $\mathcal{E}^{1^\circ}_{l, t}$ corresponds to the set edges $E_{l, t}^{1^\circ}$ complemented by their attributes and with $E_{l, t}^{1^\circ}$ being composed of the $1^\circ$ interactions at $l$, that is 
\begin{equation*}
E_{l, t}^{1^\circ} = \left\lbrace (i,j) \in E_l : i\in \mathcal{I}_{l,t}, j\notin  \mathcal{D} ^+_{ [1 : t-1]}  \text{ and } \theta^j_{t}  < l  \right\rbrace.
\end{equation*}

In addition, the set $\mathcal{V}^{1^\circ}_{l, t} $ corresponds to the vertices in $E_{l, t}^{1^\circ} $ complemented by their attributes. 

As we mentioned before, the time of infection of index cases is inferred from the observations and is defined as 

\begin{align}\label{e:law_dirac-infect}
    \Pro \left(\tau^i_I = d \; \middle| \;  \mathcal{O}^{\zeta}_t \right) = \Pro \left(\tau^i_I = d \; \middle| \;  \mathcal{O}^{\gamma, 1^\circ}_t  \right) = \delta_{\widehat{\tau} ^i_I}(d).
\end{align}

As a consequence, Equation \eqref{e:total_prob} becomes

\begin{equation*}
     \Pro \left( \bigcap_{l = t - \gamma}^t Y^{ij}_l = 0 \; \middle| \;  \mathcal{O}^{\zeta}_t \right) =  \Pro \left( \bigcap_{l = t - \gamma}^t Y^{ij}_l = 0 \; \middle| \;  \tau^i_I = \widehat{\tau} ^i _I, \mathcal{O}^{\gamma, 1^\circ}_t \right).
\end{equation*}
 
Finally,  for the   $1^\circ$CT method, the risk given  by \eqref{e:risk_2} for $j$ at $t$ is defined as  

\begin{equation*}
    R_{\gamma}^{j,1^\circ} \left( t \right) = 1 - \prod_{i \in V : i\neq j} \prod_{l=t-\gamma}^t \left(1 - \lambda ^{i\rightarrow j}_{\widehat{\tau} ^i _I,l} \mathbb{1} \left( l > \theta ^j _{t} \right) \right).
\end{equation*}

Remind that, as considered in Equation \eqref{e:cond_prob}, $\lambda ^{i\rightarrow j}_{\widehat{\tau} ^i _I,l}=0$ if $\widehat{\tau} ^i _I = \infty$.

\subsubsection{Infection risk for 2°contact tracing}\label{ssec:2-CT}

Here, we derive the computation of the risk defined by Equation \eqref{e:risk_2} by considering as possible sources of transmission both index cases and $1^\circ$ contacts. 
For any individual $j$ such that $j \notin {\mathcal D}^+_{[1:t-1]}$, when the possible source is an index case, the risk computation is analogous to the one developed for the $1^\circ$ CT method. 
On the other hand, when the possible source is a $1^\circ$ contact $i$ (such that $i \notin {\mathcal D}^+_{[1:t-1]}$), we use the parameter $\zeta$ ($\zeta \geq \gamma $) as the time-frame for the infection date of $i$, meaning that it lies in the interval of time $[t-\zeta:t-1]$. 
As a consequence, we are interested in $1^\circ$ interactions that occur in $[t-\zeta:t-1]$ 
and in the $2^\circ$ interactions that occur after a possible transmission due to a $1^\circ$ interaction in the interval of time $[t-\gamma:t]$. 

Hence, the set of observations  reduces to,

\begin{equation*}
    \mathcal{O}^{\zeta, \gamma, 2^\circ}_t = \left\lbrace \mathcal{G}_{l,t}^{1^\circ}\right\rbrace_{ l \in [t-\zeta : t]} \cup \left\lbrace \mathcal{G}_{l, t}^{2^\circ}\right\rbrace_{ l \in [t-\gamma : t]} \cup  \left\lbrace \mathcal{D} ^+_l \right\rbrace _{ l \in [1 : t-1]}  \cup \left\lbrace \mathcal{D} ^-_l \right\rbrace_{ l \in [1 : t-1]} \subset\mathcal{O}^{\zeta}_t,
\end{equation*} 
where $\mathcal{G}_{l,t}^{1^\circ} = \left( \mathcal{V}_{l, t}^{1^\circ}, \mathcal{E}^{1^\circ}_{l, t}  \right) $  is defined in Section \ref{ssec:1-CT},  $\mathcal{G}_{l, t}^{2^\circ} = \left( \mathcal{V}_{l,t}^{2^\circ}, \mathcal{E}^{2^\circ}_{l,t}  \right)  \subset \mathcal{G}_{l} $, $\mathcal{E}^{2^\circ}_{l,t}$ corresponds to the set of edges $E_{l,t}^{2^\circ}$ complemented by their attributes, and  with $E_{l, t}^{2^\circ}$  being composed of the  $2^\circ$ interactions at $l$, i.e.

$$E_{l, t}^{2^\circ} = \left\lbrace (i,j) \in E_l : \exists k\in V \text{ s.t. } (k,i) \in \bigcup_{ d \in [t-\zeta : l-1]}  E_{d,t}^{1^\circ},
j\notin  \mathcal{D} ^+_{ [1 : t-1]}  \text{ and }  \theta^j_{t} < l\right\rbrace. $$
 The set $\mathcal{V}^{2^\circ}_{l,t}$ is composed of the vertices in $E_{l,t}^{2^\circ} $ complemented by their attributes. 

Let us now consider an individual $i$ that has not been detected up to time $t-1$.
We remind that, $\theta ^ i _{t}$ is defined as the last time, prior to $t$, of a negative result test for $i$. On the other hand, $t-\zeta$ is considered as the first possible time of infection of $i$.
Hence, we denote by $\theta ^ i _{t, \zeta} = (\theta ^ i _{t} +1)\;\vee \; (t-\zeta )$ the first possible time of infection for $i$, where  the notation "$ t_1 \; \vee \; t_2 $" stands for the maximal term between $t_1$ and $t_2$.
Then,  to bring together all the possible sources of infections (index cases and others), we model the distribution of the time of infection of any individual $i\in V$, and any time $d \in [1:t-1] \cup \{+\infty\}$ as follows,
\begin{equation}\label{e:law-geo-infect}
   \Pro  \left(\tau^i_I = d \; \middle| \;  \mathcal{O}^{\zeta, \gamma, 2^\circ}_t \right) = \delta _{\widehat{\tau} ^i _I} \left( d \right) \mathbb{1} \left(i\in \mathcal{D}^+_{[1:t-1]}\right) + g^i_{t, \zeta}\left(d \right)\mathbb{1} \left(i\notin \mathcal{D}^+_{[1:t-1]} \right),
\end{equation}
where $g^i_{t, \zeta}(d)$ is the probability mass function of a truncated generalised geometric in $[\theta ^ i _{t, \zeta} : t-1]\cup \{+\infty\}$, defined as
 \begin{equation*}
   g^i_{t, \zeta}\left(d\right) = \mathbb{1}\left(\theta ^ i _{t, \zeta} \leq d \leq t-1\right) p_{d, t}^i \prod _{l=\theta ^ i _{t, \zeta} } ^{d-1} \left(1-p^i_{l, t}\right) + \mathbb{1}\left(d = \infty \right) \prod _{l=\theta ^ i _{t, \zeta} } ^{t-1} \left(1-p^i_{l, t}\right).
\end{equation*}

We denote by $p^i_{l, t}$ the probability that $i$ gets infected by some index case at time $l$, given the set of observations $\mathcal{O}^{\zeta, \gamma , 2^\circ}_t$, that is

 \begin{equation*}
   p^i_{l, t} = 1 - \prod_{k \in V: k\neq i} \Pro \left(  Y^{ki}_l = 0 \; \middle| \;  \tau^{k}_I = \widehat{ \tau}^{k}_I, \mathcal{O}^{\zeta, \gamma, 2^\circ}_t \right).
\end{equation*}

For the $2^\circ$ CT method, the risk of infection for an individual $j$ at  $t$ is defined as, 
\begin{equation*}
    R_{\gamma,  \zeta}^{j, 2^\circ } \left(  t\right) = 1 - \prod_{i \in V : i\neq j} \sum_{d \in \{1,2,...,t-1,\infty\}}  \prod_{l=t-\gamma}^t \Pro \left(  Y^{ij}_l = 0 \; \middle| \;  \tau^i_I = d, \mathcal{O}^{\zeta, \gamma, 2^\circ}_t \right)  \Pro  \left(\tau^i_I = d \; \middle| \; \mathcal{O}^{\zeta, \gamma, 2^\circ}_t \right).
\end{equation*}

\section{Simulation results}\label{sec:simulation_results}

\subsection{Simulated data}

To test our proposed method on a proper data set, we generate the data using the OpenABM-Covid19 model introduced by  \cite{hinch2021openabm}. This agent-based model simulates the spread of the COVID-19 disease on a sequence of contact networks representing the daily interactions within a population whose demographic structure is based upon UK census data. This model has several advantages: (1) it is rich enough to mimic a  dynamic social contact network at the level of a real country,  with possible large population size and a variety of individual information, in particular, the daily interactions between individuals come from three different networks depicting the contacts at home, at work and at random; (2) concerning the disease spreading, several infected statuses are available ranging from asymptomatic to pre-symptomatic statuses to mild or severe symptomatic, where the pre-symptomatic status refers to infectious individuals without symptoms; (3) it has several implementation advantages such as a very fast running time and the fact that new intervention methods, like ours, can be easily integrated into the existing code.

In the OpenABM-Covid19 model, the transmission probability takes into account the infectiousness of the source (day of infection, disease severity according to the status), the susceptibility of the recipient based on the age group, and the place where the interactions occur, putting more weight to the household interactions than the others. Indeed, the transmission probability that $i$ infects $j$ at time $t$ is defined as 

\begin{equation}\label{e:prob_abm}
\lambda_{\tau^{i}_{I}, t}^{i\rightarrow j} = 1-\text{exp}\left( -D \; f_A(a^{j})  \; f_B(b^{i}_t )\; f_C( c^{ij}_t) \int_{t - \tau^{i}_{I} - 1}^{t - \tau^{i}_{I} } f_{\Gamma}(u; \mu, \sigma ^2) du \right),
\end{equation}
where 
\begin{enumerate}
    \item $f_{\Gamma}$ accounts for the varying infectiousness over the course of the disease, and it is chosen as the density function of the Gamma distribution with mean $\mu$ and standard deviation $\sigma$,
    \item $f_B( b^{i}_t)$ is the severity of the individual $i$ considered as a possible source at time $t$ ($i$ can be susceptible, asymptomatic, pre-mild, mild, pre-severe, severe or removed),
    \item $f_A(a^j)$ is the relative susceptibility of the recipient $j$, which depends on the age group of $j$, and is normalised by the mean number of daily interactions by age group,
    \item $f_C(c^{ij}_t)$ is the strength of the interaction (if it is at home, work or random) between $i$ and $j$ at $t$,
    \item $D$ scales the overall infection rate.
\end{enumerate}

For more details on the OpenABM-Covid19 model, and in particular, on the functions $f_A$, $f_B$ and $f_C$, the interested reader can refer to Supplementary Material and 
\cite{hinch2021openabm}

 \subsection{Intervention strategy}

\begin{figure}[ht!]
\centering
\includegraphics[width=0.99\textwidth]{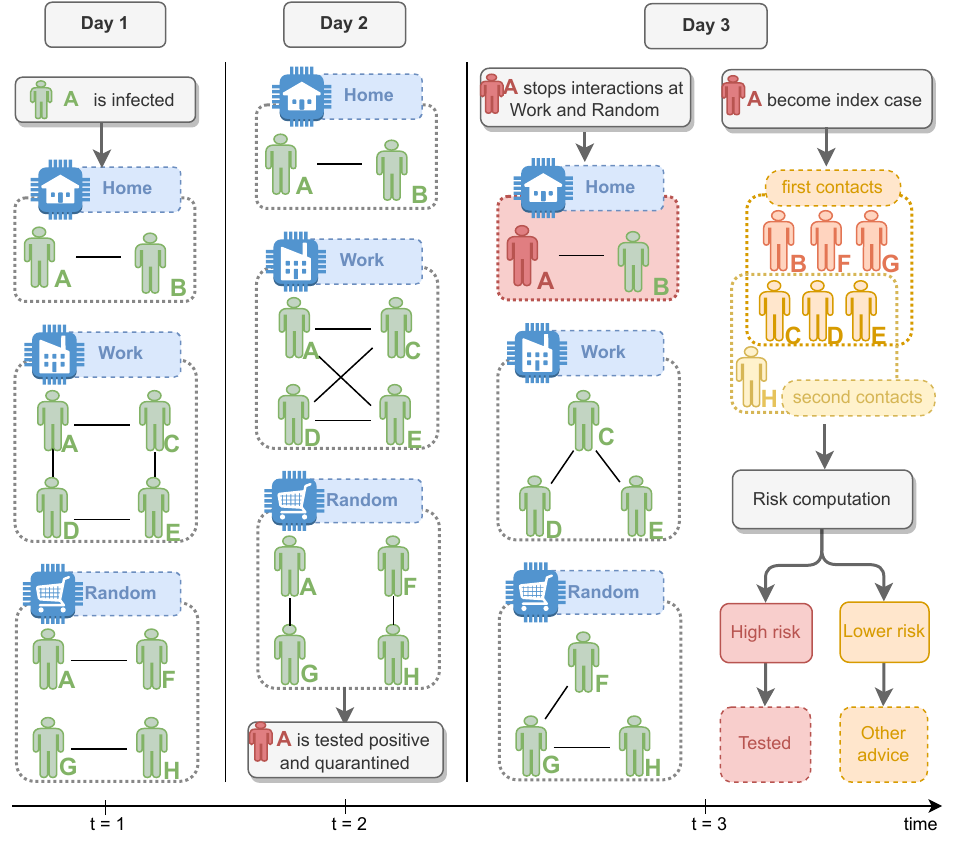}    
\caption{Interactions between individuals during 3 days. When the individual A is detected, A is quarantined and her/his time of infection is estimated. The $2^\circ$contact tracing method trace forward the first and second contacts in interaction with A after the estimated time of infection, and compute the risk for those individuals. The individuals with the highest risk are tested.}
\label{fig_2-CT}
\end{figure}

The simulation starts at time $t=0$  with $N$ individuals. At the beginning, all individuals are susceptible ($S$), except for a small number $N_0$ of infectious individuals (“patients zero”). 
Every day, starting from $t=1$, a proportion $p_s$, respectively $p_{m}$, of individuals with newly developed severe and mild symptoms are 
tested, detected and 
quarantined.
Later, at a fixed date $t_0$ in $[1:T]$, the intervention based on the risk calculation presented in Section \ref{ssec:risk} starts, and it is carried out daily until the mitigation of the epidemic or the end of the study. At any $t\geq t_0$, the intervention strategy based on the $2^\circ$CT method consists of tracing $1^\circ$ and $2^\circ$contacts, computing their risk of infection, and ranking them according to their risk values. See Figure \ref{fig_2-CT} for an illustration of how the proposed $2^\circ$CT method works for a simple scenario of three days. Then, the first $\eta$ individuals in the ranking are tested, and the newly detected ones become index cases and are quarantined. The default quarantine protocol stops the interactions in the occupation and random network, but those within the household are maintained. 
On a given day, it may happen that the number of traced individuals is smaller than $\eta$, in which case we randomly select and test individuals among those who have not been detected, until reaching the number of $\eta$ daily available tests; for those detected, we set their time of infection at $\gamma$ days prior to their detection (since we do not have information to infer their time of infection from previous contacts with detected individuals). 
As already mentioned, the tests are assumed to be perfect, and we suppose that the test results are available the same day on which the tests are performed. 

Notice that the $1^\circ$CT method is a particular case of the $2^\circ$CT method, and hence, the intervention related to the $1^\circ$CT method  is analogous to the one for the $2^\circ$CT method, except that only the $1^\circ$contacts are traced and ranked.

To compute the risk of infection for the $1^\circ$CT and $2^\circ$CT methods, we use the transmission probability considered  in \cite{hinch2021openabm} and defined by Equation \eqref{e:prob_abm}, with 
$\widehat{b}^i_t$ and $\widehat{\tau}^i_I$ in place of the true values $b^i_t$ and $\tau^i_I$. Due to the way the  infectiousness is modelled, we have that $\lambda ^{i \rightarrow j} _ {\tau, t} \approx 0$ if $t-\tau \geq 15$. The latter, combined with a  significant gain of the computational cost of our method, leads us to reduce the set of index cases as follows 

\begin{align*}
    \mathcal{I}_{l, t} = \left\lbrace i\in V : l \in ]\widehat{\tau}^i_{I}\vee (t-21) : \widehat{\tau}^i_{R}[ \right\rbrace.
\end{align*}

\subsection{Results}

In this section, we present the results obtained using the intervention based on the risk, through the simulation of different scenarios.
For all the simulations, the propagation of the epidemic is identical until $t_0$, while it might change after $t_0$, when the intervention method starts, depending on the particular scenario. In the figures presented later, each thin line represents the result obtained for the realisation associated with one \textit{seed}, while the thick lines correspond to the average of all the realisations.

\subsubsection{Estimation of the time of infection}\label{ssec:estimation_time_infection}

As seen in Sections \ref{ssec:1-CT} and \ref{ssec:2-CT}, 
the estimated time of infection of index cases has a direct impact and an indirect impact on the computation of the risk of infection since (1)  the probability of transmission depends on it and (2) the selection of the risky interactions relies on it. 
The estimation of the time of infection is computed on the day of detection of the individual, and it remains constant over time after this day. Remind that, for any individual $j$ who has not been detected, we have set $\widehat{\tau}^{j}_{I}=\infty$.
 
At any time $t\ge t_0$, the individuals are tested because of their symptoms or because they have been traced and selected based on their risk of infection. Among the symptomatic individuals, let us denote by $m$ ($m\in \N$) the expected number of days it takes to develop symptoms after the day of infection. In our simulations, we used $m=6$ as in  \cite{hinch2021openabm}. 

For any individual $j$ detected at $t$, i.e. $j \in \mathcal{D}_t^+$, we define $\widehat{\tau}^{j}_{I}$ as follows, 
\begin{align*}
\widehat{\tau}^{j}_{I} = 
\begin{cases} 
(t - m) \vee  (\theta^j_t+1) 
&  \text{if} \ j \ \text{is detected by symptoms at $t$, }  \\
\widehat{\tau}^{j, 1^\circ}_{I}\mathbb{1}\left(\widehat{\tau}^{j, 1^\circ}_{I} \neq \infty \right)\;  
+ \;  \widehat{\tau}^{j, 2^\circ}_{I} \mathbb{1}\left( \widehat{\tau}^{j, 1^\circ}_{I} = \infty \right)  &  \text{if} \ j \ \text{is detected by  risk at } t,
\end{cases} 
\end{align*}
where $\widehat{\tau}^{j, 1^\circ}_{I}$ and  $\widehat{\tau}^{j, 2^\circ}_{I}$ 
are defined respectively as  the minimal time of the $1^\circ$  and $2^\circ$ interactions of $j$, lying within the time-frame $[t-\gamma:t]$, that is
\begin{align*}
\widehat{\tau}^{j, 1^\circ}_{I} & = \min \left\lbrace  d\in [t-\gamma : t] : \exists  i \in V \text{ s.t. } (i,j) \in E_{d, t}^{1^\circ} \right\rbrace, \\
\widehat{\tau}^{j, 2^\circ}_{I} & = \min \left\lbrace  d\in [t-\gamma : t] : \exists  i \in V \text{ s.t. } (i,j) \in E_{d, t}^{2^\circ} \right\rbrace,
\end{align*}
where,  by convention we set $\min \emptyset = \infty$.

In order to evaluate the effectiveness of the estimator $\widehat{\tau}^{j}_{I}$ in the mitigation strategy, we propose another estimator $\widehat{\alpha}^{j}_{I}$, which is constant for all index cases provided that they do not have a previous negative test result, that is
\begin{align*}
\widehat{\alpha}^{j}_{I} = 
\begin{cases} 
(t - m) \vee  (\theta^j_t+1) 
&  \text{if} \ j \ \text{is detected by symptoms at $t$,}  \\
(t - \gamma\;) \vee  (\theta^j_t+1) 
&  \text{if} \ j \ \text{is detected by risk at $t$. }
\end{cases} 
\end{align*}

\begin{figure}[h!]
  \centering
  \begin{subfigure}[b]{0.244\textwidth}
         \centering
         \includegraphics[width= \textwidth]{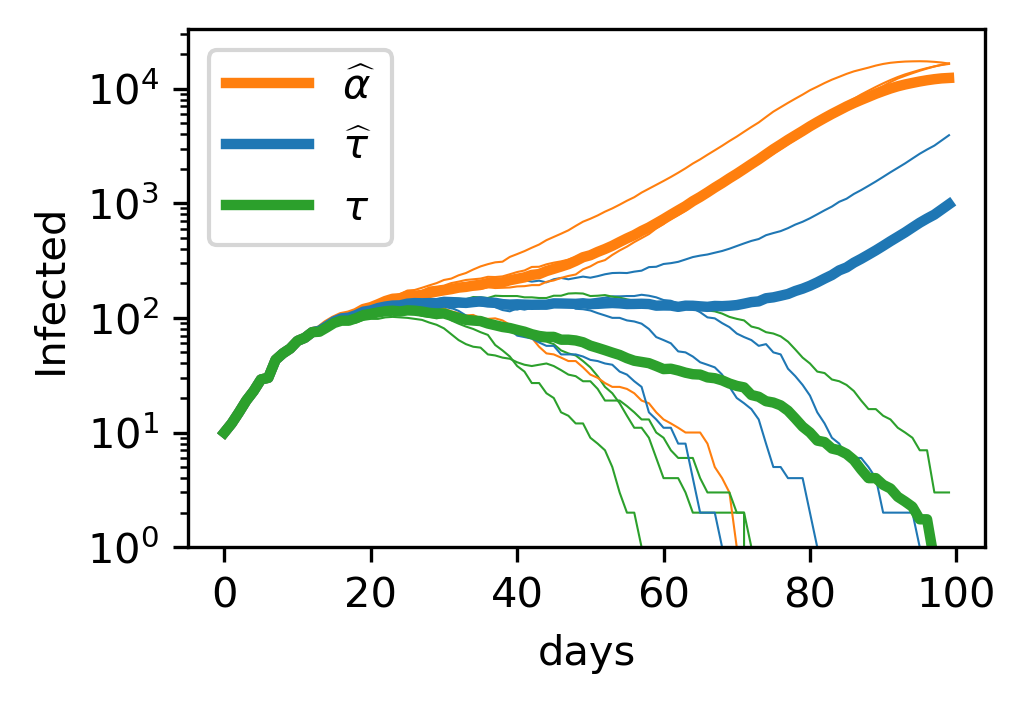 }
         \caption{$1^\circ$CT($\gamma = 6$)}
  \end{subfigure}
  \begin{subfigure}[b]{0.244\textwidth}
         \centering
         \includegraphics[width=\textwidth]{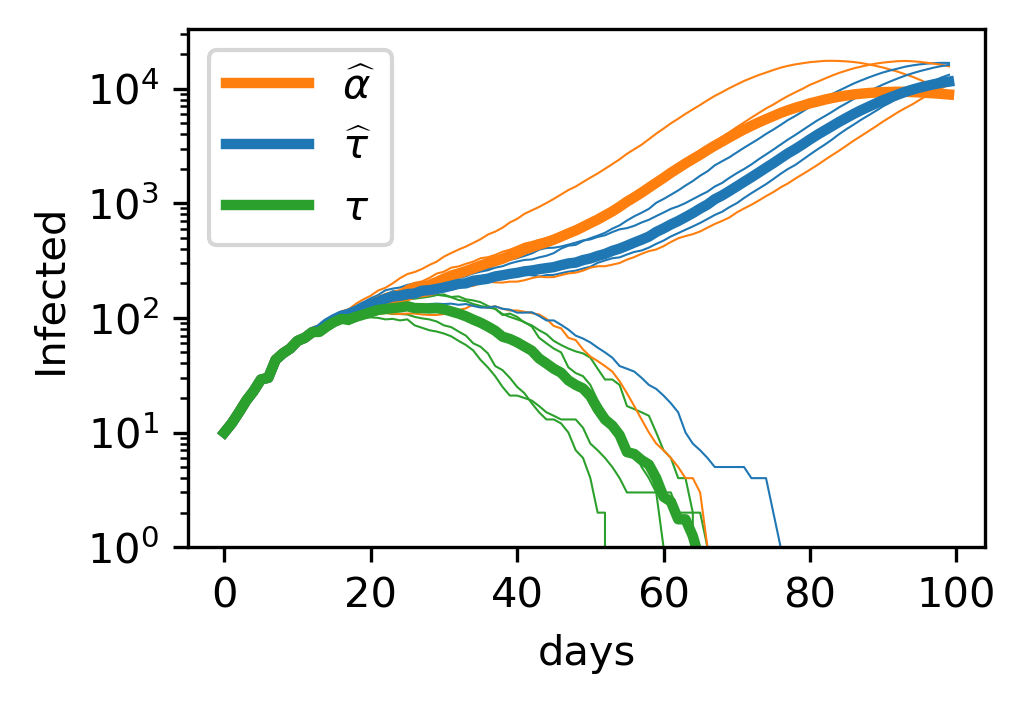} 
         \caption{$2^\circ$CT($\gamma = 6$, $\zeta = 7$)}
  \end{subfigure}
   \begin{subfigure}[b]{0.244\textwidth}
         \centering
         \includegraphics[width= \textwidth]{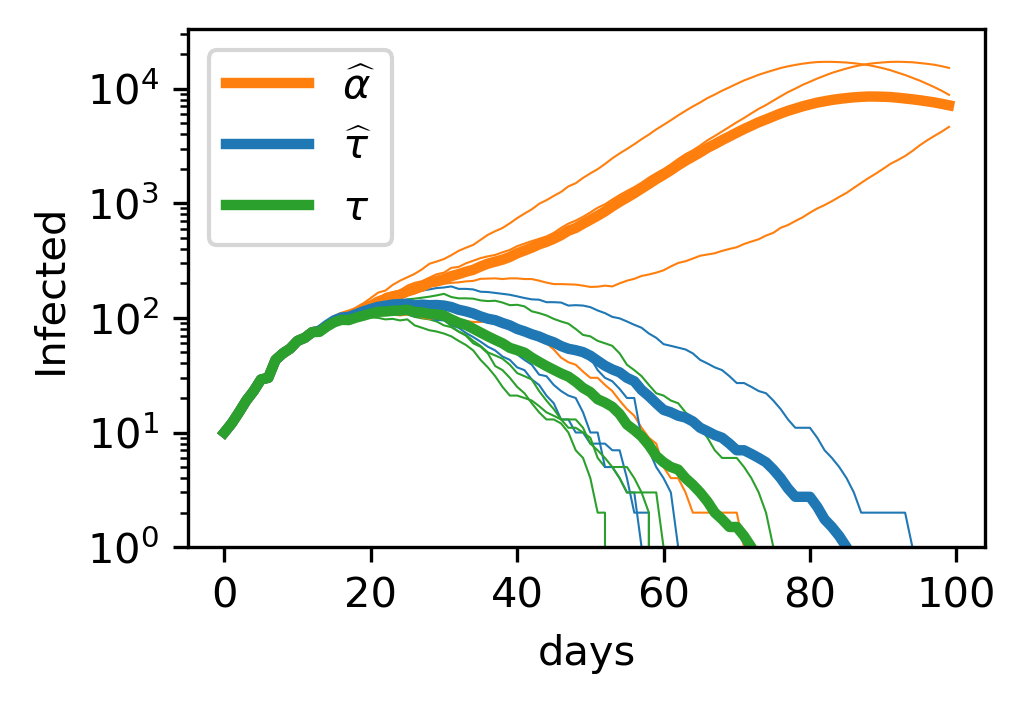}         
         \caption{$2^\circ$CT($\gamma = 6$, $\zeta = 8$)}
         \end{subfigure}
    \begin{subfigure}[b]{0.244\textwidth}
         \centering
         \includegraphics[width= \textwidth]{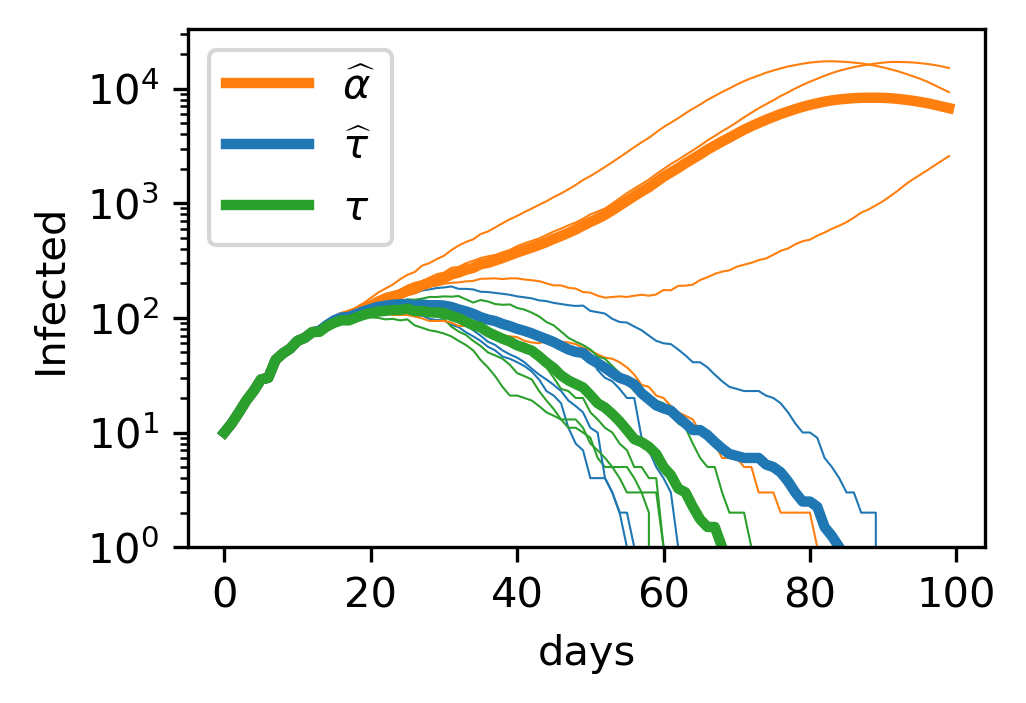}         
         \caption{$2^\circ$CT($\gamma = 6$, $\zeta = 9$)}
         \end{subfigure}
\caption{Effect of the estimators $\widehat{\tau}_I^i $ (blue), $\widehat{\alpha}_I^i$ (orange) and the real time of infection $\tau_I^i $ (green) for the $1^\circ$CT and $2^\circ$CT methods on the spread of the epidemic. The figures display the number of infectious individuals during $T = 100$ days among a population of size $N=50K$, when the intervention starts at day $t_0 = 12$, with $N_0=10$ patients zero, $\eta = 125$ daily available tests, a proportion $p_s=1$ of detected severe symptomatic individuals, and a proportion $p_m = 0.75$ of detected mild symptomatic individuals.}
\label{estimation_time_infection}
\end{figure}

\begin{figure}[h!]
\centering
  \hspace{-0.25cm}
    \includegraphics[width=0.5 \textwidth]{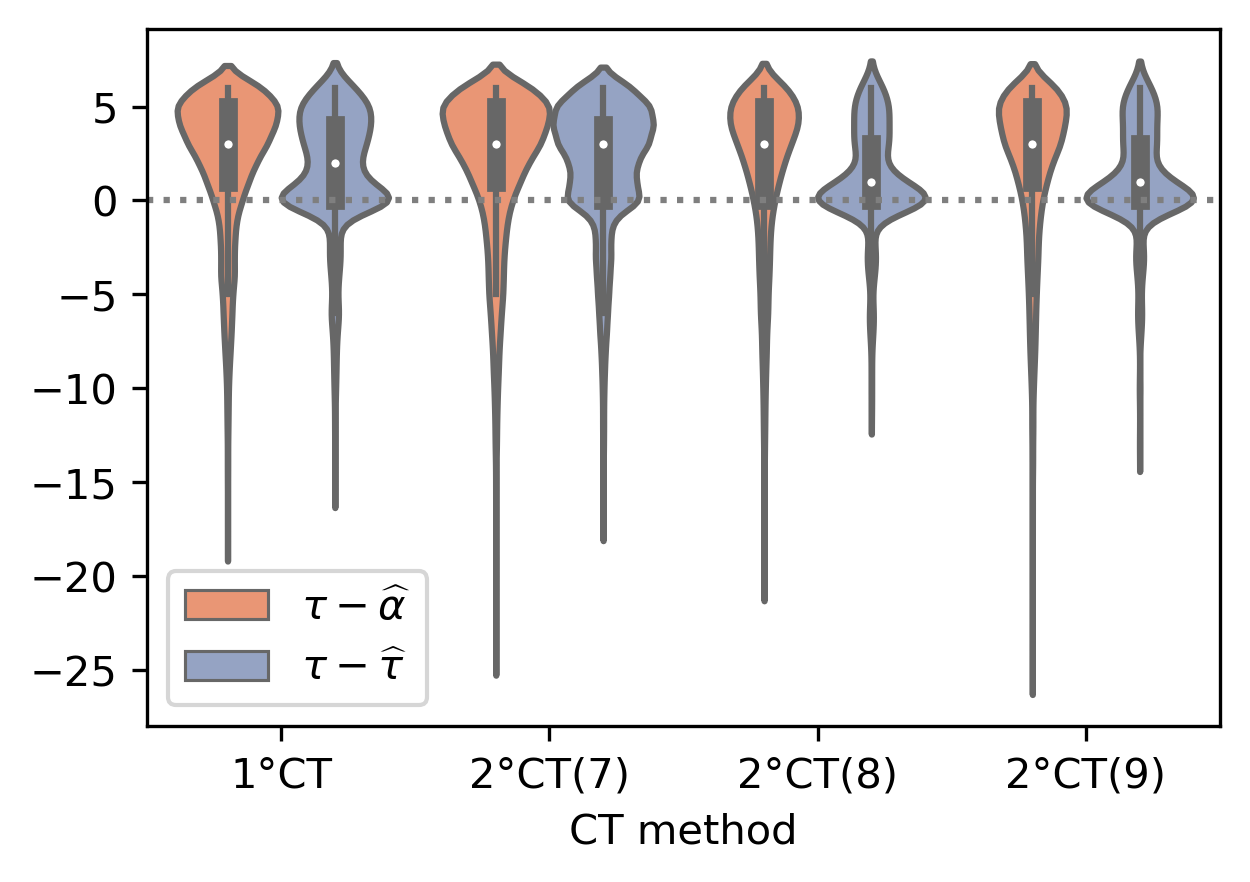}
\caption{Box-plots of the differences  $\tau_I^i - \widehat{\tau}_I^i$ (orange) and $\tau_I^i - \widehat{\alpha}_I^i$ (blue) for the individuals detected by risk in
the $1^\circ$CT and $2^\circ$CT methods with the range of parameter values  $\zeta = 7, 8, 9$, and $\gamma = 6$. Each box-plot has been built from four seeds, with $T=100$, $N=50K$, $t_0=12$, $N_0=10$, $\eta=125$, $p_S=1$, and $p_m=0.75$. }
\label{box_plot}
\end{figure}

To compare the effect of $\widehat{\tau}_I^i$, $\widehat{\alpha}_I^i$, and $\tau_I^i$ (the real infection time of $i$), on the mitigation of the epidemic, we simulate the same intervention strategy with these three different times of infection and for different sets of parameter values. We depict in Figure~\ref{estimation_time_infection} the number of infectious individuals in logarithmic scale through time, and for the same simulated trajectories, we display in Figure \ref{box_plot} the box-plots of the empirical distribution of the differences $\tau_I^i - \widehat{\alpha}_I^i$ and  $\tau_I^i - \widehat{\tau}_I^i $. 
 
The results in Figure \ref{estimation_time_infection} show that, for a broad range of parameter values, the use of the estimated infection time $\widehat{\tau}_I^i$ is more effective in the mitigation of the epidemic than the constant estimator $\widehat{\alpha}_I^i$. This improvement is more pronounced in panels (c) and (d), in which $\widehat{\tau}_I^i$ provides results that are almost as good as the ones obtained with $\tau_I^i$, the true time of infection.
If we take a look at the corresponding box-plots in Figure \ref{box_plot}, we can see that the empirical distribution of $\tau_I^i - \widehat{\tau}_I^i $ is more concentrated around the empirical median, and hence, has less variability, than the one of $\tau_I^i - \widehat{\alpha}_I^i$. Moreover, for $\zeta =8, 9$, the empirical median of $\tau_I^i - \widehat{\tau}_I^i $ is much closer to zero than the one of $\tau_I^i - \widehat{\alpha}_I^i$. In these latter cases, the good performance of the estimator $\widehat{\tau}_I^i$ allows reaching the mitigation of the epidemic faster than the constant estimator (see panels  (c)-(d)  in Figure \ref{estimation_time_infection}). 

\subsubsection{Time-frame for the time of infection }

In a context of limited resources, providing an effective strategy for mitigating an epidemic goes through the detection at an early stage of the most likely infected individuals. Indeed, the detection of the latter before they become highly infectious is preferable. 
Hence, tuning the parameter $\gamma$, which corresponds to the considered infection time-frame for individuals at risk,  plays a key role in providing an effective strategy for the mitigation of the epidemic. 
There is a trade-off between large and small values of $\gamma$. For large values of $\gamma$, one expects to detect more individuals since, by construction, the set of observations increases with time. On the other hand, small values of $\gamma$ allow us to concentrate the efforts on the more recently infected individuals, before they propagate the disease, and discard those individuals that were infected a long time ago.

\begin{figure}[ht]
     \centering
     \begin{subfigure}[b]{0.3\textwidth}
         \centering
         \includegraphics[width=\textwidth]{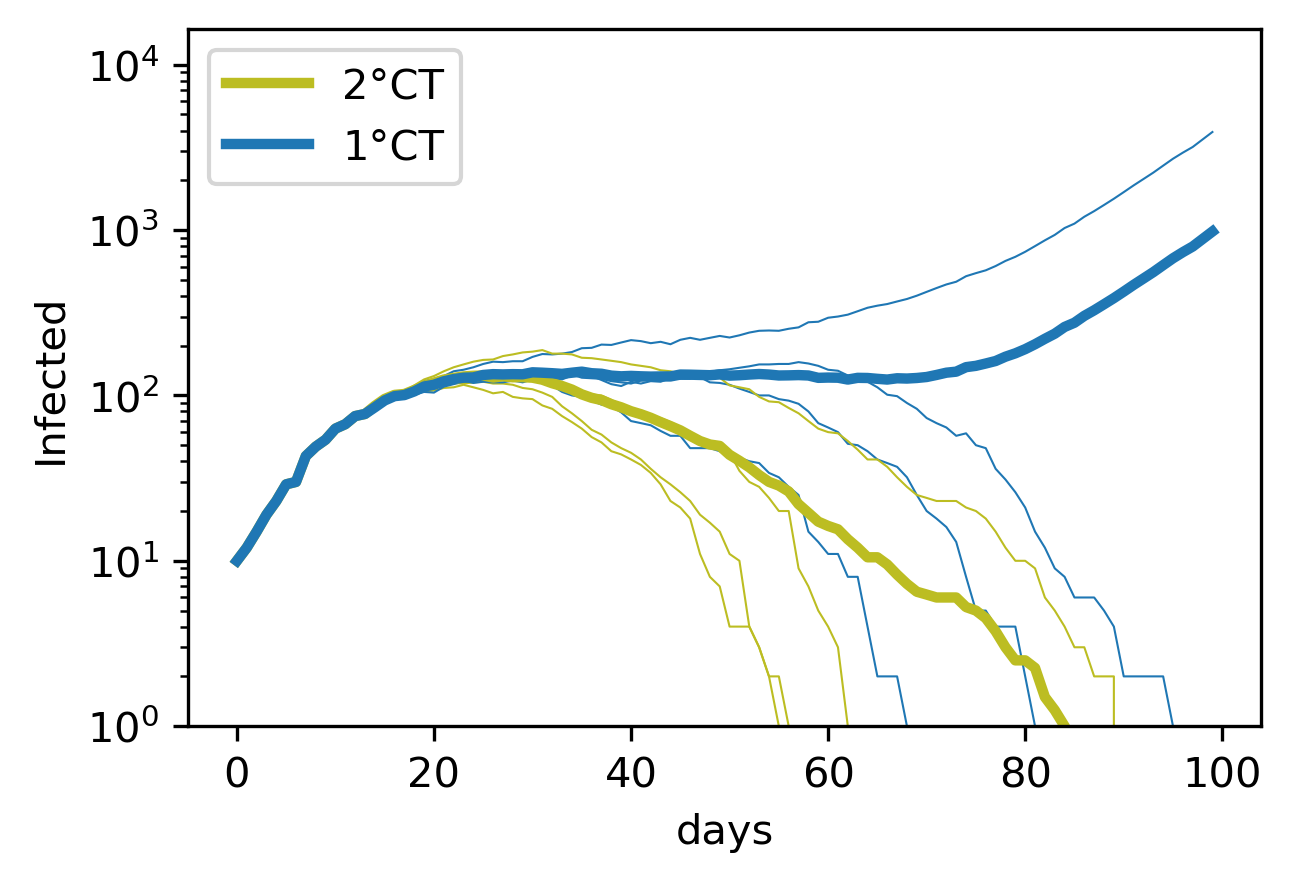}
         \caption{$\gamma = 6$}
         \label{fig:y equals x}
     \end{subfigure}
     \hfill
     \begin{subfigure}[b]{0.3\textwidth}
         \centering
         \includegraphics[width=\textwidth]{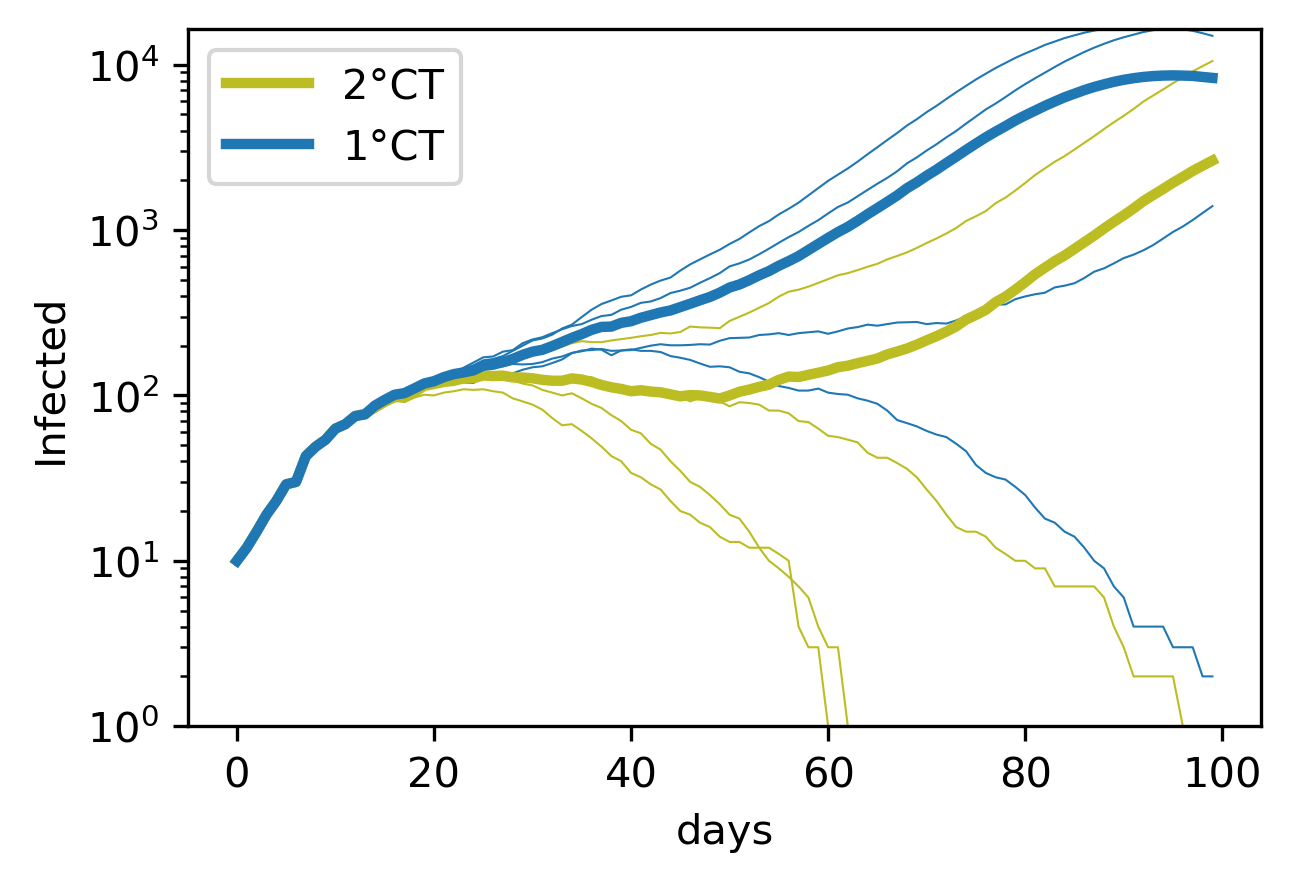}
         \caption{$\gamma = 10$}
         \label{fig:three sin x}
     \end{subfigure}
     \hfill
     \begin{subfigure}[b]{0.3\textwidth}
         \centering
         \includegraphics[width=\textwidth]{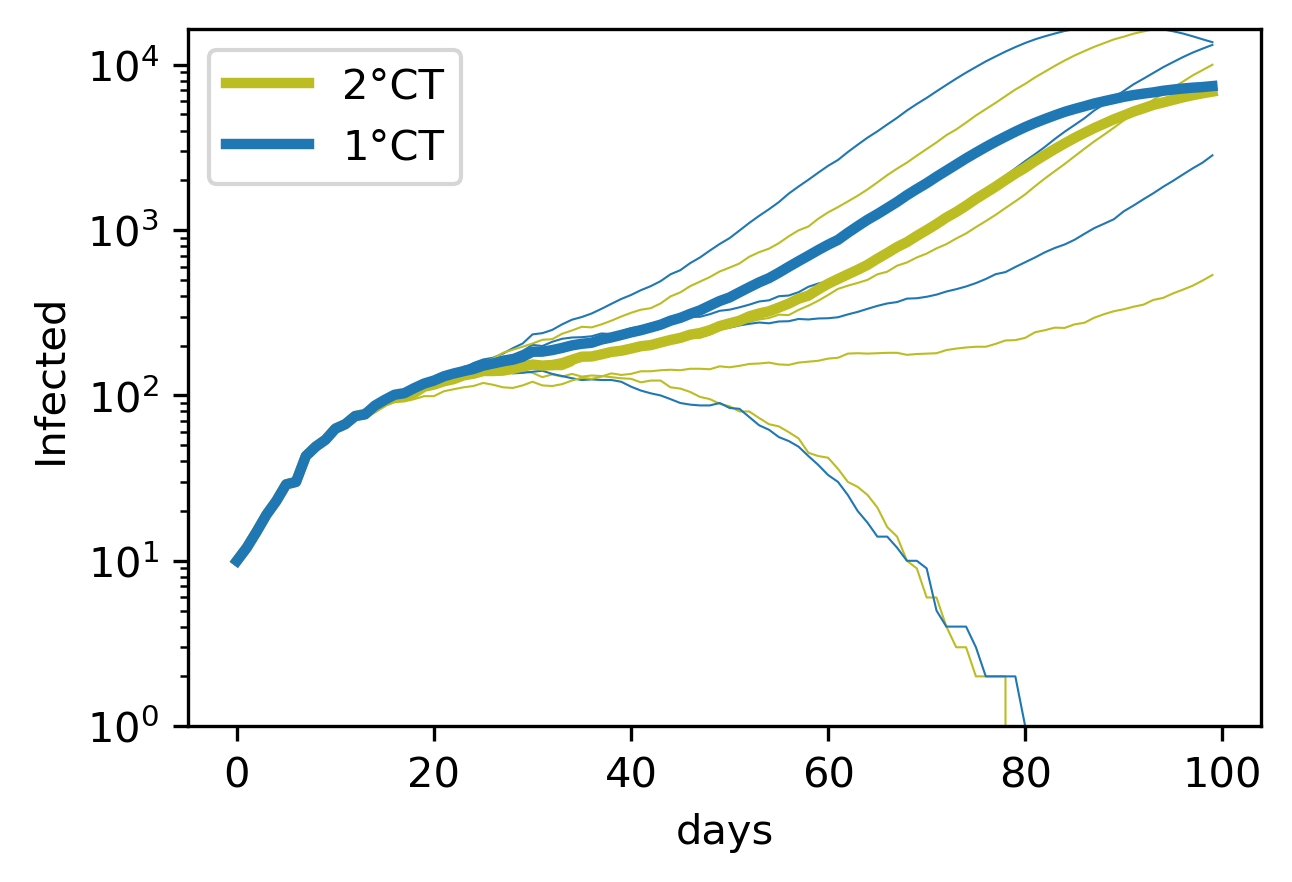}
         \caption{$\gamma = 14$}
         \label{fig:five over x}
     \end{subfigure}
        \caption{Effect of $\gamma$ on the epidemic spreading for strategies $1^\circ$CT (blue) and $2^\circ$CT (yellow) when $\zeta = \gamma + 3$.
Each plot has been built from four seeds, with $T=100$, $N=50K$, $t_0=12$, $N_0=10$, $\eta=125$, $p_S=1$ and $p_m=0.75$. }
\label{fig_comparison_gamma}
\end{figure}

\begin{figure}[h!]
\centering
    \includegraphics[width=0.5 \textwidth]{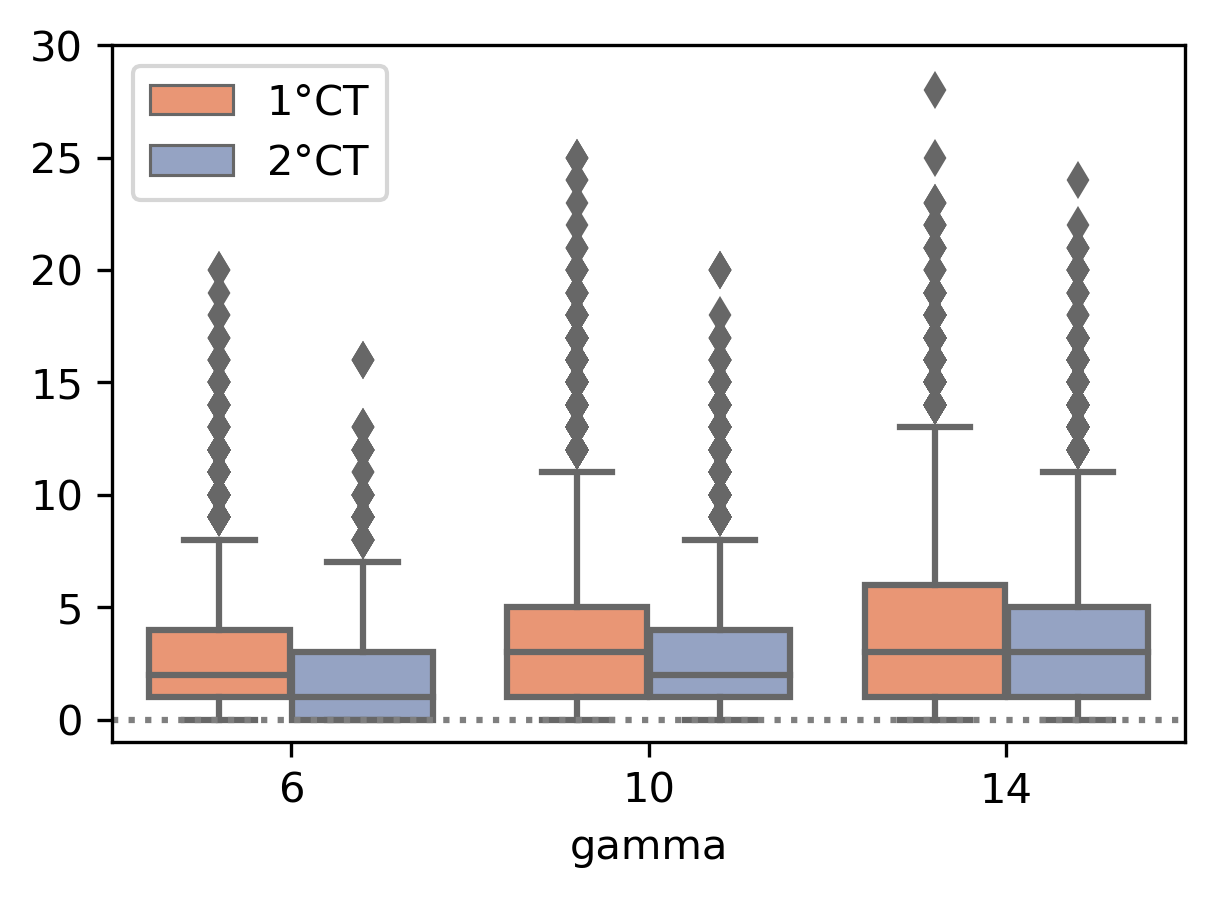}
\caption{Box-plots of the differences between the time of infection and the time of detection for the individuals detected by risk in the $1^\circ$CT and $2^\circ$CT methods with the parameters values  $\gamma = 6, 10, 14$, and for $2^\circ$CT $ \zeta = \gamma +3$, where $T=100$, $N=50K$, $t_0=12$, $N_0=10$, $\eta=125$, $p_S=1$, $p_m=0.75$ for four different seeds. }
\label{box_plot_time_detection}
\end{figure}

In Figure~\ref{fig_comparison_gamma}, we study the impact of the values of $\gamma$ on the mitigation of the epidemic for $1^\circ$CT and $2^\circ$CT methods, showing the number of infectious individuals in logarithmic scale through time. For these simulations we keep $\zeta$ constant with respect to $\gamma$ (the choice of the value of the parameter $\zeta$ is further discussed in Section \ref{ssec:comp}). In Figure~\ref{box_plot_time_detection}, we focus on the effect of $\gamma$ on the early or late detection of the individuals by displaying the box-plots of the difference between the time of detection and the time of infection for the individuals detected by risk. Each box-plot in Figure~\ref{box_plot_time_detection} has been built from the same simulations presented in the Figure~\ref{fig_comparison_gamma}.

Figure~\ref{box_plot_time_detection}  shows that, for both methods,  the detection of infected individuals is faster with a small value of $\gamma$; this enables a faster progression of the epidemic mitigation as seen in Figure  \ref{fig_comparison_gamma}. From both figures, we conclude that taking $\gamma = 6$ in the two methods makes them effective in detecting and quarantining individuals at an early stage of infection and improves the allocation of resources. 

\subsubsection{Probability of transmission}

The probability of transmission plays also a role in the ability of the contact tracing strategy to mitigate the epidemic. For the $1^\circ$CT and $2^\circ$CT methods, the probability of transmission is involved in the risk of infection calculated at $t$ for an individual $j$ and depends directly on the estimation of the time of infection (see   Equation \eqref{e:prob_abm}). 
 
\begin{figure}[ht]
     \centering
     \begin{subfigure}[b]{0.49\textwidth}
         \centering
         \includegraphics[width=\textwidth]{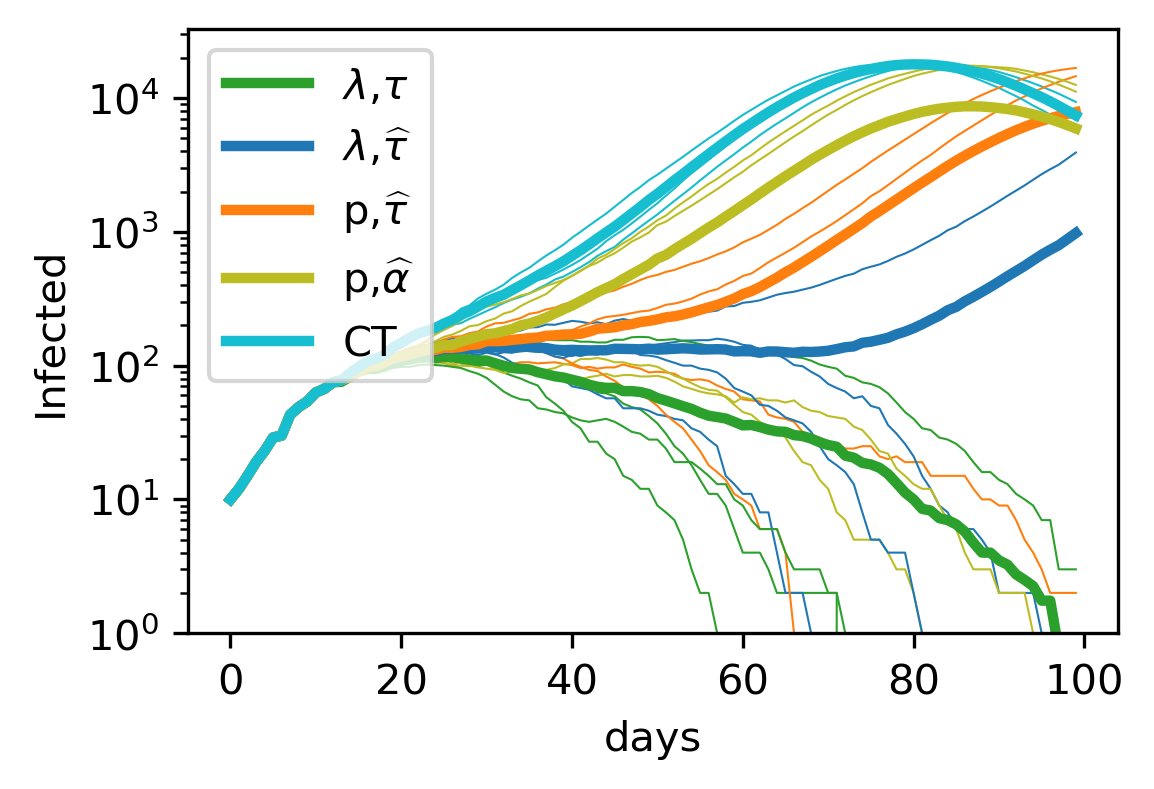}
         \caption{$1^\circ$CT($\gamma = 6$)}
         \label{prob_1-CT}
     \end{subfigure}
     \hfill
     \begin{subfigure}[b]{0.49\textwidth}
         \centering
         \includegraphics[width=\textwidth]{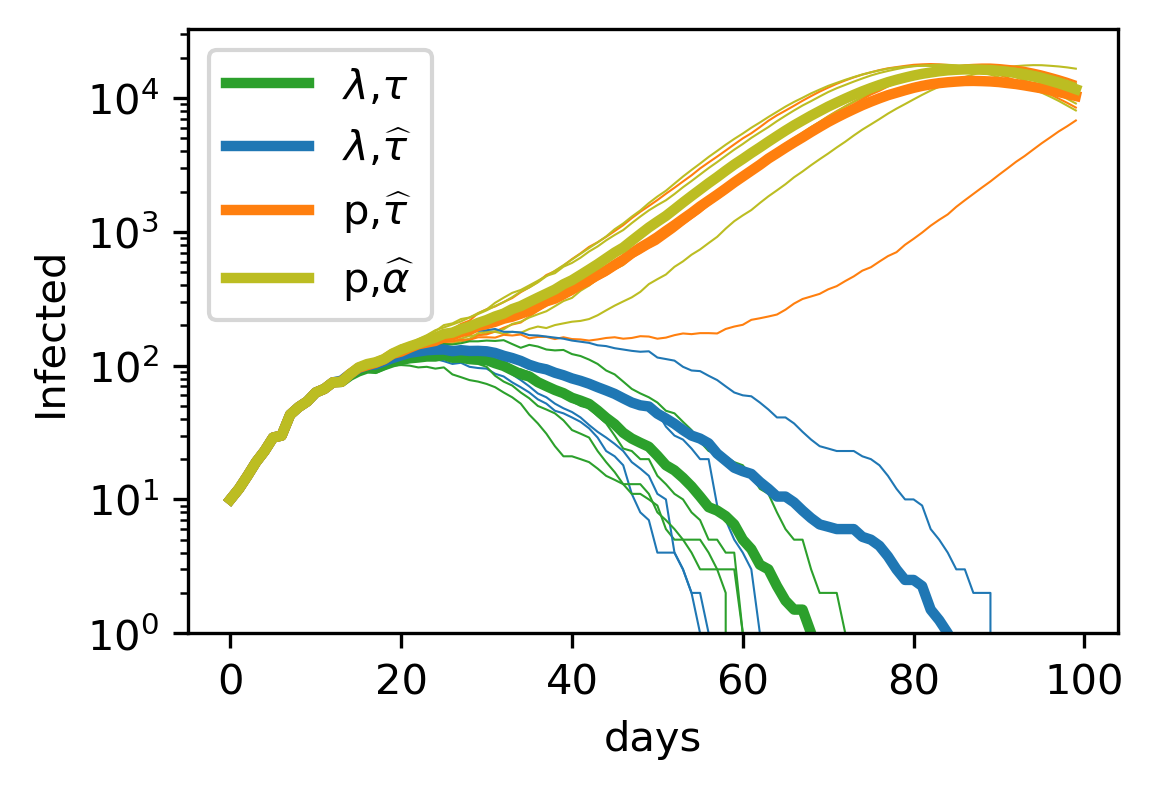}
         \caption{$2^\circ$CT($\gamma = 6$, $\zeta = 9$)}
         \label{prob_2-CT}
     \end{subfigure}
        \caption{Effect of the transmission probability function on the spreading of the epidemic for $1^\circ$CT and $2^\circ$CT methods. 
Each plot has been done with $T=100$, $N=50K$, $t_0=12$, $N_0=10$, $\eta = 125$, $p_m=0.75$, $p_S=1$, $\gamma = 6$ and $\zeta = 9$.}
       \label{fig_comparison_prob_function}
\end{figure}

 Here we compare the $1^\circ$CT and $2^\circ$CT methods for a range of values of the  pair ``probability of transmission and real or estimated time of infection'', i.e., for 
 $(\lambda_{\tau_I^i}^{i \rightarrow j}, {\tau_I^i})$, 
 $(\lambda_{\hat{\tau}_I^i}^{i \rightarrow j}, {\hat{\tau}_I^i})$,
 $(p,  {\hat{\tau}_I^i})$
   and $(p, {\hat{\alpha
  }_I^i})$, 
 where $\lambda_{\hat{\tau}_I^i}^{i \rightarrow j}$ is defined by Equation \eqref{e:prob_abm},  $\hat{\tau}_I^i$, $\hat{\alpha}_I^i$ are defined in Section \ref{ssec:estimation_time_infection}, $\tau_I^i$ is the true infection time of $i$    and $p=1/2$  is a constant. 
 For all of these pairs, Figure \ref{fig_comparison_prob_function} depicts the number of infectious individuals in logarithmic scale through time, for $\gamma=6$ and $\zeta=9$.

Both panels in Figure \ref{fig_comparison_prob_function}  show that the results given by the pair $(p,  {\hat{\tau}_I^i})$ (in orange) are much better than the ones obtained with $(p,  {\hat{\alpha}_I^i})$ (in yellow), confirming that it is crucial to accurately estimate the infection time.
Moreover, the probability of transmission that depends on the individual and interaction attributes (in dark blue) considerably improves the results compared with the choice of a constant probability of transmission. 

In  Figure \ref{prob_1-CT}, the results provided by the  CT method are included; the CT method consists of ranking individuals according to their number of interactions with detected individuals in the time-frame $[t-\gamma: t]$. It is important to highlight that the CT method differs from the $1^\circ$CT method with $(p, {\hat{\alpha}_I^i})$, since in the CT method the dates of the negative test results are not taken into account. Figure \ref{prob_1-CT} shows that including the information on negative test results has a positive effect on the mitigation of the epidemic.
 
\subsubsection{Comparison with other ranking methods} \label{ssec:comp}

To evaluate the efficiency of the proposed $1^\circ$CT and $2^\circ$CT methods to  mitigate an epidemic, we compare them with three other ranking strategies:

\begin{enumerate}
\item  Random Selecting (RS): individuals not previously detected are ranked randomly.
\item Contact Tracing (CT): individuals not previously detected are ranked according to their number of interactions with detected individuals in the time-frame $[t-\gamma: t]$.
\item Mean-Field (MF):  individuals not previously detected are ranked according to the mean-field risk approximation presented in  \cite{baker2021epidemic}.
\end{enumerate} 

The mean-field procedure depends on two parameters: (1) $\rho_{MF}$, the mean time elapsed between the time of infection and the time of detection and (2) $t_{MF}$, a parameter called \textit{integration time} on the MF method, meaning that given new observations, the probabilities are updated in the interval $[t-t_{MF}: t]$. For the following simulations we consider $\rho_{MF} = 5$ and $t_{MF} = 10$, as in  \cite{baker2021epidemic}. In Section 2 of the Supplementary Material, we provide further details about the main differences in the computation of the mean-field risk and the $2^\circ$ CT risk.

\begin{figure}[h!]
     \centering
     \begin{subfigure}[b]{0.3\textwidth}
         \centering
         \includegraphics[width=\textwidth]{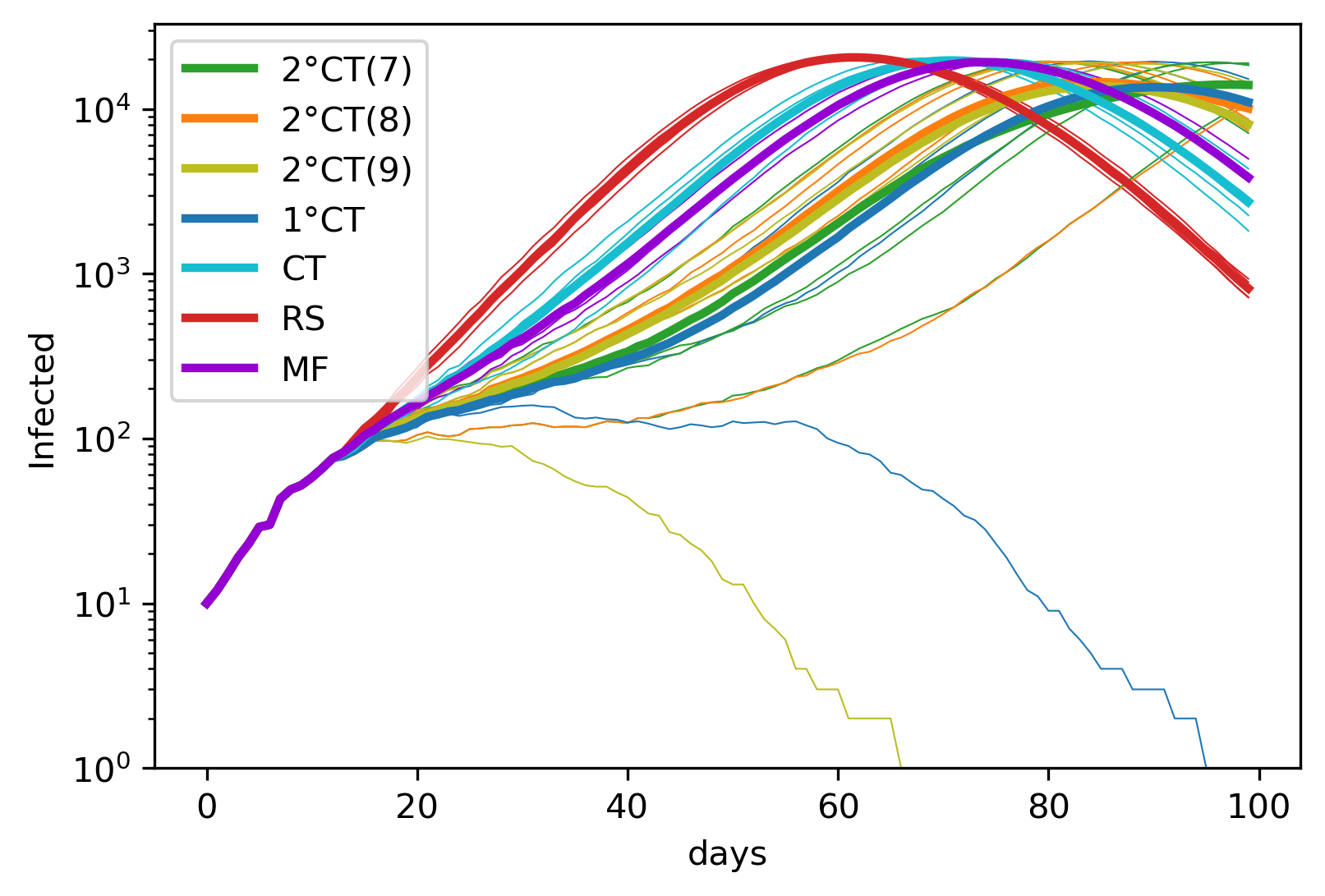}
         \caption{$\eta = 125, p_m = 0.5$}
     \end{subfigure}
     \hfill
    \begin{subfigure}[b]{0.3\textwidth}
         \centering
         \includegraphics[width=\textwidth]{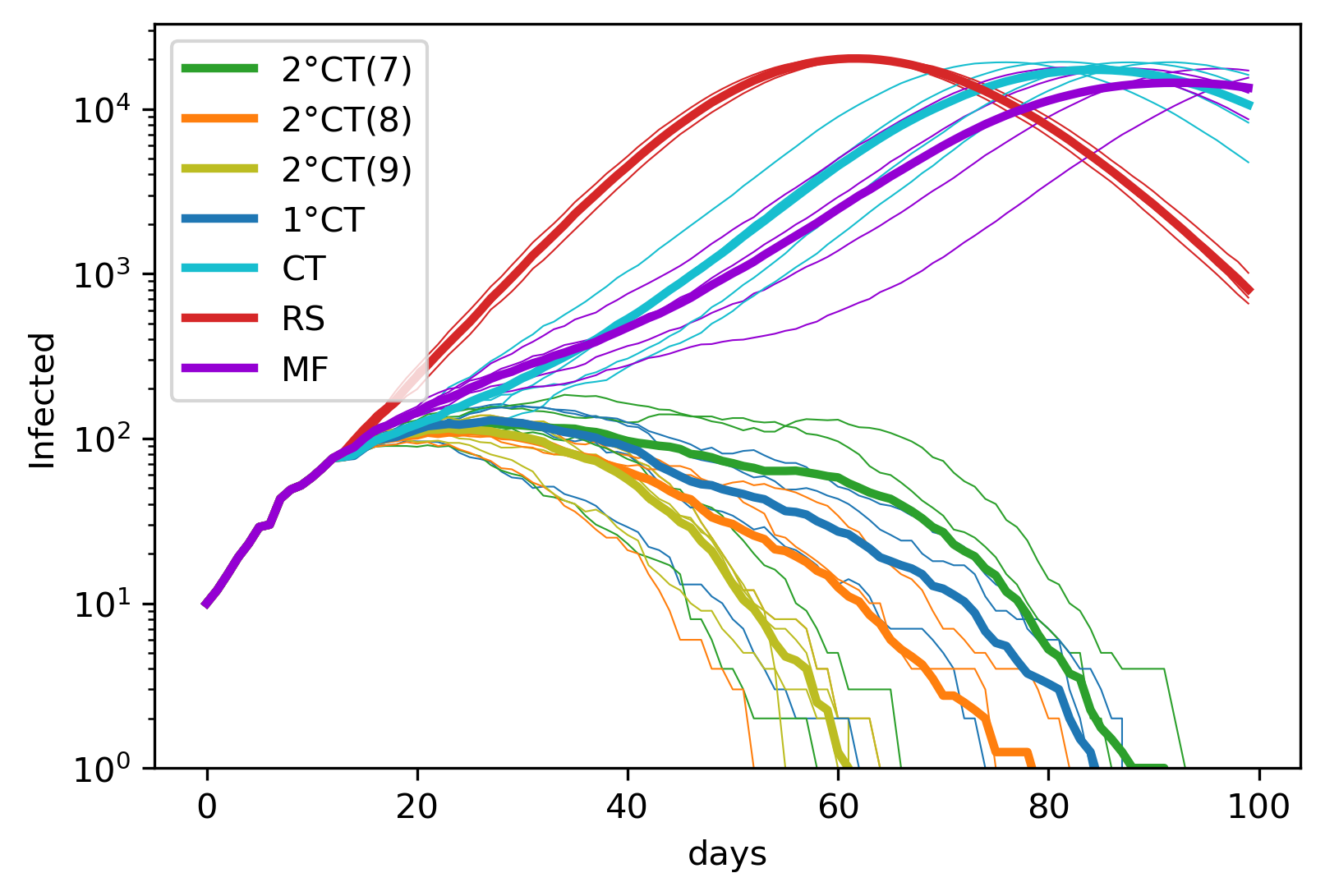}
         \caption{$\eta = 250, p_m = 0.5$}
     \end{subfigure} 
       \hfill
    \begin{subfigure}[b]{0.3\textwidth}
         \centering
         \includegraphics[width=\textwidth]{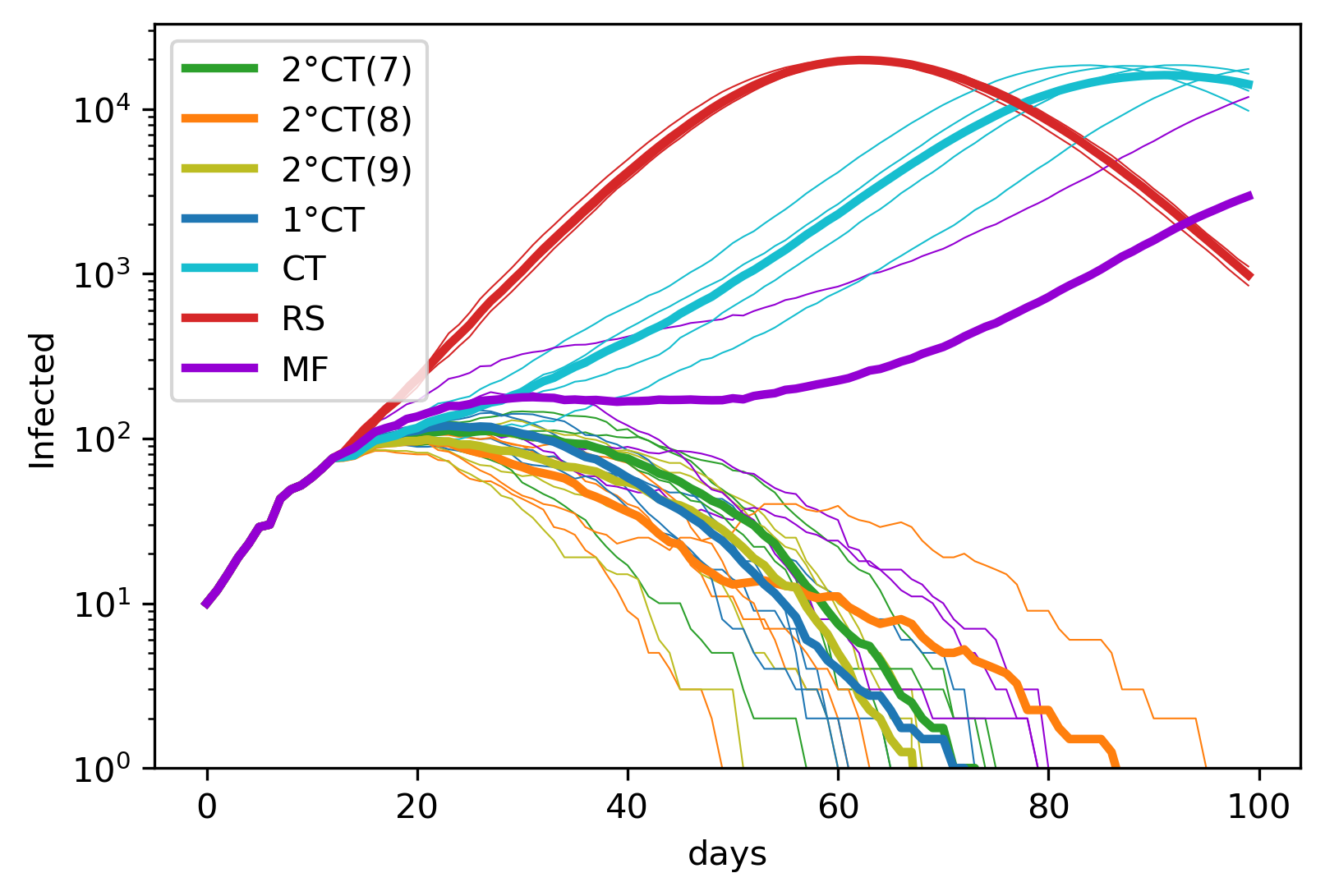}
         \caption{$\eta = 400, p_m = 0.5$}
     \end{subfigure} 
     \hfill
     \begin{subfigure}[b]{0.3\textwidth}
         \centering
         \includegraphics[width=\textwidth]{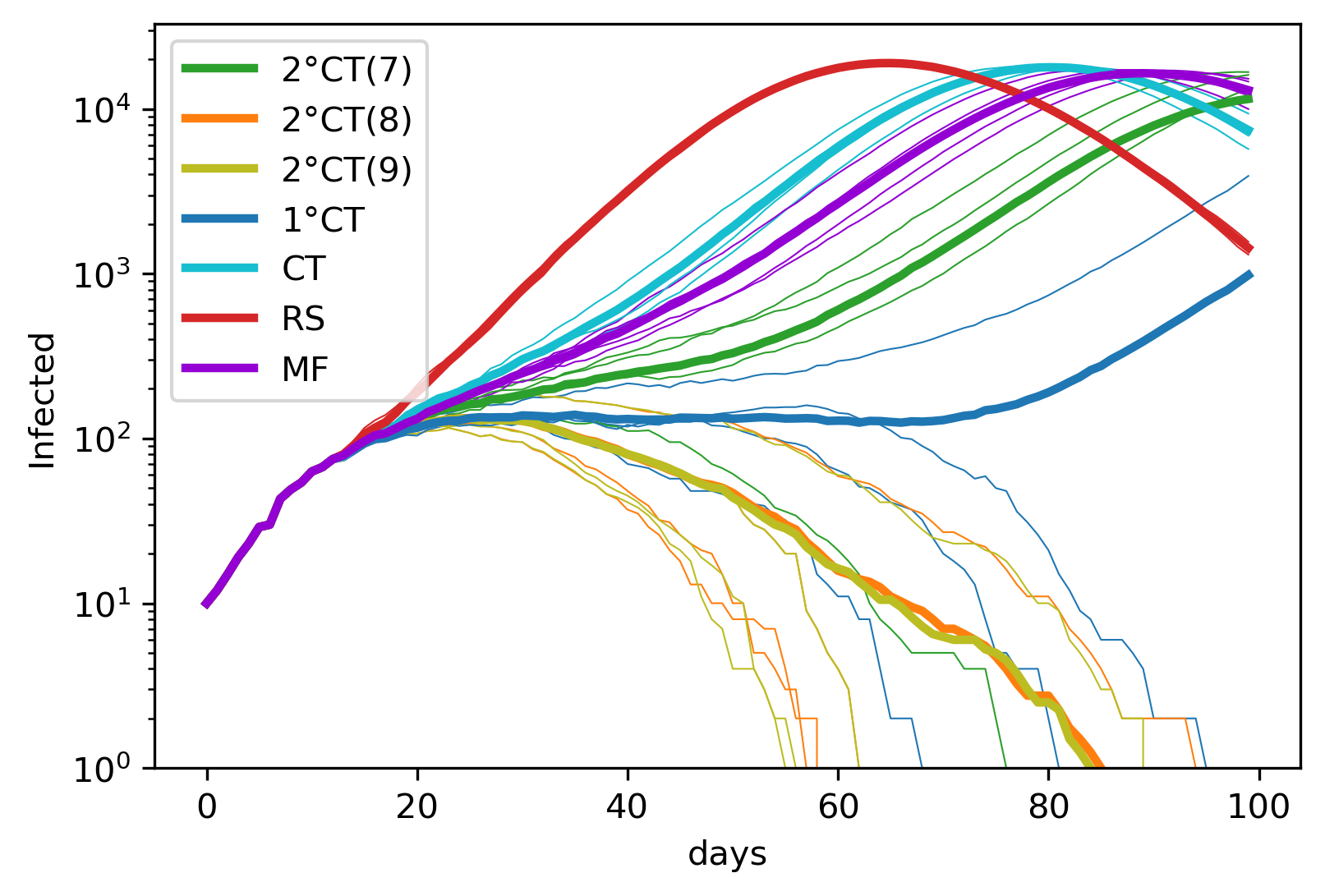}
         \caption{$\eta = 125, p_m = 0.75$}
     \end{subfigure}
     \hfill
    \begin{subfigure}[b]{0.3\textwidth}
         \centering
         \includegraphics[width=\textwidth]{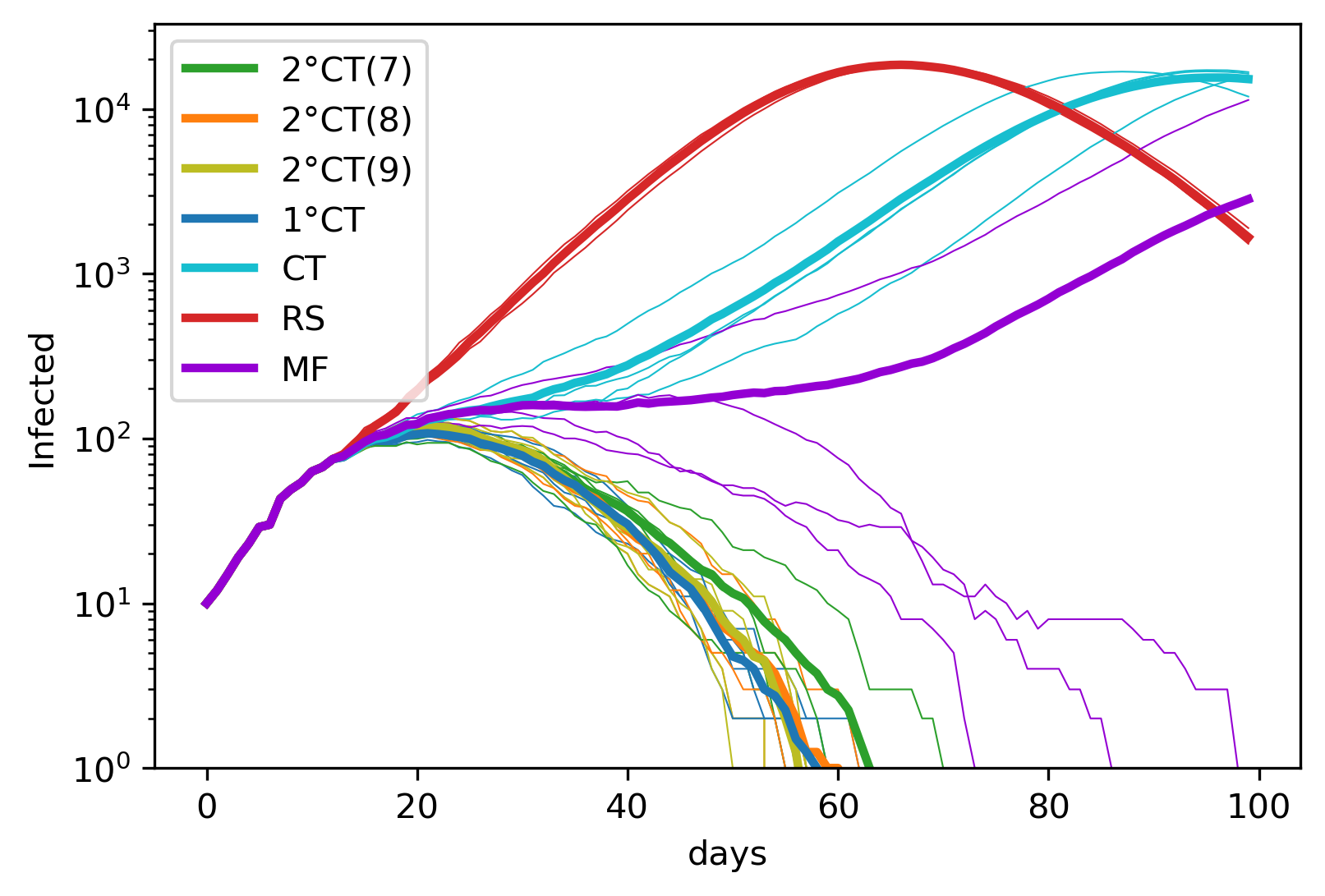}
         \caption{$\eta = 250, p_m = 0.75$}
     \end{subfigure} 
       \hfill
    \begin{subfigure}[b]{0.3\textwidth}
         \centering
         \includegraphics[width=\textwidth]{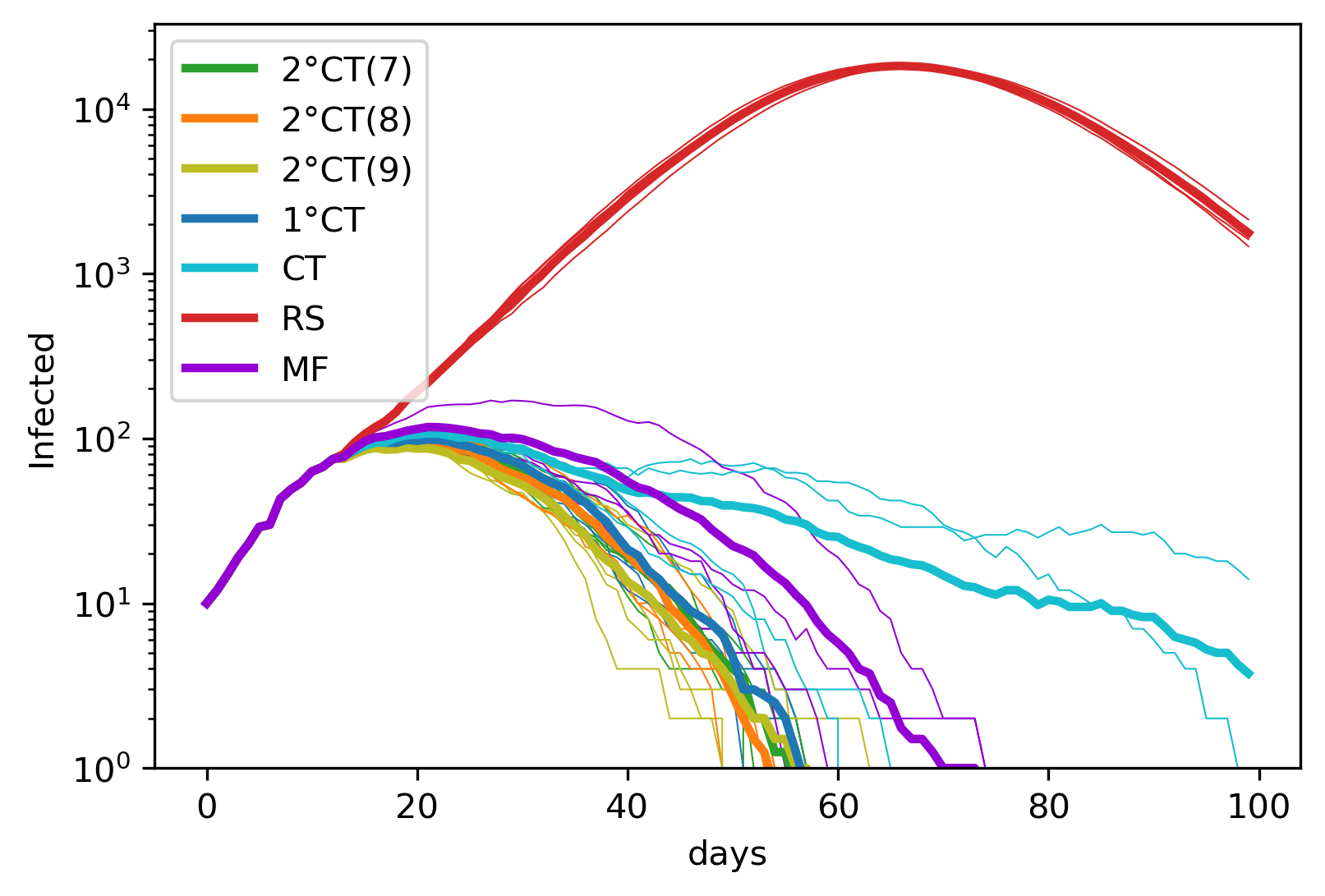}
         \caption{$\eta = 400, p_m = 0.75$}
     \end{subfigure} 
        \caption{Effect of the parameters $\eta$ (the number of daily available tests, increasing from left to right) and $p_m$ (the proportion of daily detected individuals with mild symptoms, increasing from top to bottom) on the epidemic spreading for the strategies $1^\circ$CT, $2^\circ$CT,  CT, RS and MF.   
In all simulations we consider $T=100$, $N=50K$, $t_0=12$, $N_0=10$ and $p_S=1$. 
The estimation of the infection time for the $1^\circ$CT and $2^\circ$CT  strategies is given by   $\widehat{\tau}_I^i$. 
We fix the parameters $\gamma = 6$ and $\zeta = 7, 8, 9$ (indicated in the legend as $2^\circ$CT(7), $2^\circ$CT(8), $2^\circ$CT(9), respectively). The values for the  parameters in  MF strategy  are $\rho_{MF} = 5$ and $t_{MF} = 10$.}
\label{fig_comparison_eta_mild}
\end{figure}

We compare the five strategies in Figure~\ref{fig_comparison_eta_mild}, in which we display the number of infectious individuals in logarithmic scale through time across a broad range of values for the parameters. In particular, we increase the number of daily available tests from the left panels to the right ones, and we increase the proportion of daily detected mild symptomatic individuals from top to bottom.
As expected for all strategies, a higher value of  $p_m$ and/or $\eta$ improves the mitigation of the epidemic in terms of the duration and the total number of infected individuals. The simulations show that our proposed methods ($1^\circ$CT and $2^\circ$CT) improve considerably the results compared to the MF and the usual CT, which are all better than the RS strategy. The latter does not mitigate the epidemic even with a high number of daily available tests and a high value of $p_m$, while the MF and CT methods achieve the mitigation for a large value of $\eta$. 
We also study the $2^\circ$CT method for different time-frames in which the $1^{\circ}$contact can get infected, that is in $[t-\zeta: t]$, where we consider $\zeta = 7$ in green, $\zeta = 8$ in orange and $\zeta= 9$ in yellow.
Figure \ref{fig_comparison_eta_mild} shows that the results are improved as $\zeta$ increases. In particular, it should be noticed that the $2^\circ$CT method with $\zeta = 8$ and $\zeta = 9$ gives better results than the $1^\circ$CT method.  However, the $1^\circ$CT method requires less individual information and therefore it is better in terms of privacy restrictions. From these results, a trade-off can arise between 
getting better results with computationally demanding ($2^\circ$CT method) and preserving individual privacy with simpler and faster computation. Indeed, it is worth mentioning that for a high enough number of daily available tests and/or a high enough proportion of mild observed, the methods $1^\circ$CT and $2^\circ$CT have similar effects on the mitigation of the epidemic; hence in this case,  we recommend the use of the $1^\circ$CT method than the $2^\circ$CT method.

\section{Discussion}\label{discussion}

In this paper, we have introduced a method for the computation of the probability of infection of individuals in interaction, based on forward contact tracing, that considers at risk not only the direct contacts of detected individuals but also their subsequent contacts. We have called our method second-degree contact tracing ($2^\circ$CT). The proposed method consists of estimating the individual infection risk by considering all possible chains of transmission up to second-degree contacts, coming from index cases. We propose a mitigation strategy consisting of using the risk approximation to rank individuals and allocate the (limited) number of daily available tests according to this ranking. We have evaluated interventions based on our risk ranking through simulations of a fairly realistic agent-based model calibrated for COVID-19 epidemic outbreak (the Oxford OpenABM-Covid19 model). We have considered different scenarios to study the role of key quantities such as the number of daily available tests, the contact tracing time-window,
the transmission probability per contact (constant versus depending on multiple factors), and the age since infection (for varying infectiousness). We found that, when there is a limited number of daily tests available, our method is capable of mitigating the propagation more efficiently than random selection, than the usual contact tracing (ranking according to the number of contacts with detected individuals), and than some other approaches in the recent literature on the subject. 
Additionally, our risk computation method can be easily adapted to the mitigation of other transmissible diseases spreading on contact networks.

One of the main difficulties in many forward contact tracing approaches for transmission diseases such as COVID-19, is to know the time of infection of the detected individuals. This quantity is in general not observed since in most cases individuals ignore from whom and when they got infected. However, inferring the time of infection is necessary for at least two reasons: (1) to know from which date the contacts of the detected individual should be traced, (2) to accurately assess the risk of infection in the case of varying transmissibility during the course of the disease. 
Given these arguments, we have considered age-dependent infectiousness, meaning that the probability of transmission from a source $i$ depends on the time since infection of $i$. Moreover, we have proposed an efficient estimation of the time of infection for detected individuals, which is more accurate than considering a constant infection time, as it is often proposed in the literature. The results show how this estimation improves the contact tracing method in terms of the number of infectious individuals through time. Indeed, it allows to achieve in some cases almost as good results as considering the real date of infection. 

Our $2^\circ$CT method encompasses the first degree contact tracing method, called here $1^\circ$CT method.
We have found that with a limited number of available tests, the $2^\circ$CT method is more effective in the mitigation of an epidemic than the $1^\circ$CT method. However, with a large enough number of available daily tests, both $1^\circ$CT and $2^\circ$CT methods provide similar results; in this case, we recommend the use of the $1^\circ$CT method because of a simpler and faster computation and better preserving individual privacy. Besides our results show that the proposed $1^\circ$CT and $2^\circ$CT methods can be very effective compared with the usual contact tracing, the mean-field risk approximation or the random selection of individuals to test.

Despite its advantages, our intervention method has some limitations. Firstly, the tests are assumed to be perfect, meanwhile in the literature \cite{baker2021epidemic}, \cite{herbrich2020crisp} and \cite{2023notimetowaste} consider the sensitivity and specificity of the tests. Another limitation of our method is that we run our simulations for a population of 50k individuals. In \cite{baker2021epidemic} the population size is 500k, in \cite{guttal2020risk} it is up to 100k,  while in \cite{herbrich2020crisp} and \cite{2023notimetowaste} it is limited to 10k. Another assumption that should be relaxed in a future work is that we do not include uncertainty in the list of contacts of the traced individuals, and it would be interesting to analyse how this could affect the efficacy of the intervention.

\section*{Data availability}

In this work, we use simulated data coming from a realistic agent-based model calibrated for the COVID-19 epidemic outbreak, see \cite{hinch2021openabm}. The code of this model can be found here \href{https://github.com/BDI-pathogens/OpenABM-Covid19}{https://github.com/BDI-pathogens/OpenABM-Covid19}.

\section*{Code availability}

The repository with the code is available in 
\href{https://github.com/gbayolo26/risk_estimation}{https://github.com/gbayolo26/risk_estimation}.

\bibliographystyle{rss}
\bibliography{references}

\newpage
\pagebreak
\clearpage
\begin{center}
\normalfont{\Large\sffamily\bfseries{Supplementary material for "Test allocation based on risk of infection from first and second order contact tracing"}}
\end{center}

\setcounter{equation}{0}
\setcounter{figure}{0}
\setcounter{table}{0}
\setcounter{page}{1}
\setcounter{section}{0}
\makeatletter

\section{OpenABM-Covid19 model}

The OpenABM-Covid19 model is a realistic model for the simulation of the COVID-19 epidemic propagation, see   \cite{hinch2021openabm}. It mimics the interactions between individuals taking into account patterns from UK population. In this model the interactions between individuals are represented by networks that changes every day, this networks are characterised by three different sub-graphs describing the interactions at home, workplace and random. The disease spreads through the networks considering the epidemiological characteristics of COVID-19. Thus, the model takes into account asymptomatic infections and different status of symptomatic infections including stages of severity. Individuals can develop mild or severe symptoms and start by being in a pre-symptomatic state, in which they are infectious but have no symptoms. Once an individual is recovered the model allows immunity.

For the simulations, we use the version presented in \href{https://github.com/aleingrosso/OpenABM-Covid19}{https://github.com/aleingrosso/OpenABM-Covid19}, in which our test allocation method and tracing techniques can be efficiently integrated. In the following section, we explain this model in details, in particular the generation process of the individual interactions and the disease transmission, as well as the parameters presented in this version and in consequence in our simulations.

\subsection{Interactions }
  
In the OpenABM-Covid19 model the undirected graph $\mathcal{G}_t  = \left( \mathcal{V}, \mathcal{E}_t  \right) $ have the following characteristics,

\begin{enumerate}
\item $\mathcal{V} = \{ (i, a^i) : i \in V \text{ and } a^i \in A \}$ is constant through time and it describes individuals $i$ and their associated  age group $(a^i)$, 
with $V = \{1,...,N\}$ and $A = \{A_1, A_2 \ldots,A_{9}\} $ the set of age-groups ordered by decade from “ $0-9$ years" to “ $80+$ years ".

\item  $\mathcal{E}_t = \{ (i,j,c^{ij}_t) : (i,j) \in E_t \text{ and } c^{ij}_t \in C \}$ is the set of the edges describing the interactions at time $t$ between the corresponding individuals, supplemented by the place in which this interaction occurs. Then, $C = H \cup O \cup U $, with $H = \{H_1, H_2,\ldots,H_{K}\} $ the set of $K$ distinct households according to UK demographics data;  $O = \{O_1, \ldots,O_{5}\} $ the set of distinct occupation activities (primary school, secondary school, workers, retired and elderly); and $U$ the random interactions (interactions at supermarket, public transport, etc). Every individual is assigned to a single household and a single occupation place. The construction of the membership to each household in $H$ and each occupation place in $O$, as well as the mean number of daily interactions that individuals have, is dependent on the individual age-group. 

 For any $t  \geq 0$, if $i$ and $j$ are in interaction at time $t$, it exists a unique edge $(i,j)\in E_t$. In particular, it is assumed the following:
\begin{enumerate}
\item if two individuals $i$ and $j$ are in the same \textbf{household} $H_m$ for $m \in \{1, 2, \ldots, K \}$, then they are in interaction at all times, i.e.,

$$\Pro \left( (i,j)\in E_t, c^{ij}_t = H_m  \; \middle| \;  i \in H_m, j\in H_m  \right) = 1, \quad \forall t \geq 0,$$

and if the two individuals are in different households, then they never interact in the household network, i.e.,

$$\Pro \left( (i,j)\in E_t, c^{ij}_t \in H  \; \middle| \;  i \in H_m, j\notin H_m  \right) = 0, \quad \forall t \geq 0;$$

\item 
for any $m=1,\ldots,5$, a fixed sub-graph $G_m$ represents 
the \textbf{occupations} network 
 associated with the occupation activity $O_m$; the $G_m$'s are modelled as small-world networks. If two individuals $i$ and $j$ are in interaction in the sub-graph $G_m$ for $m \in \{1, 2, \ldots, 5 \}$, then at time $t$ each connection is activated with probability $p = 1/2$ , i. e., 

$$\Pro \left( (i,j)\in E_t , c^{ij}_t = O_m \; \middle| \; (i,j) \in G_m \right) = 1/2, \quad \forall t \geq 0, $$

and if the two individuals are not in interaction in $G_m$ for $m \in \{1, 2, \ldots, 5 \}$, then they never interact in the occupation network $O_m$, i.e.,  

$$\Pro \left( (i,j)\in E_t, c^{ij}_t = O_m \; \middle| \;  (i,j)\notin G_m \right) = 0, \quad \forall t \geq 0.$$

\item 
the \textbf{random} interactions are selected  uniformly over all the remaining possible interactions, in particular excluding previous connections. 

\end{enumerate}  
 
\end{enumerate}

\subsection{Transmissions}

Recall that in the OpenABM-Covid19 model, the probability  that $i$ infects $j$ at $t$ is defined as follows,

\begin{equation*}
\lambda_{\tau^{i}_{I}, t}^{i\rightarrow j} = 1-\text{exp}\left( -D \; f_A(a^{j})  \; f_B(b^{i}_t )\; f_C( c^{ij}_t) \int_{t - \tau^{i}_{I}-1}^{t - \tau^{i}_{I}} f_{\Gamma}(u; \mu, \sigma ^2) du \right),
\end{equation*}

where    $f_{\Gamma}$ is  the Gamma density function with mean $\mu$ ($\mu = 6$) and standard deviation $\sigma$ ($\sigma = 2.5)$. The parameter $D$ scales the overall infection rate ($D = 5.75$). The function $f_A(a^j)$ represents the susceptibility of the recipient $j$  given her/his age group and is determined by,

\begin{equation*} 
f_A(a^j) = 
    \begin{cases}       
        & 0.71/\overline{I}_1, \quad \text{ if } a^j \in A_1  \\
        & 0.74/\overline{I}_2, \quad \text{ if } a^j \in A_2  \\
        & 0.79/\overline{I}_3, \quad \text{ if } a^j \in A_3  \\
        & 0.87/\overline{I}_4, \quad \text{ if } a^j \in A_4  \\
        & 0.98/\overline{I}_5, \quad \text{ if } a^j \in A_5  \\
        & 1.11/\overline{I}_6, \quad \text{ if } a^j \in A_6  \\
        & 1.26/\overline{I}_7, \quad \text{ if } a^j \in A_7  \\
        & 1.45/\overline{I}_8, \quad \text{ if } a^j \in A_8  \\
        & 1.66/\overline{I}_9, \quad \text{ if } a^j \in A_9  
            
    \end{cases}
\end{equation*}

where $\overline{I}_k$ is the mean number of daily interactions in the k-th age-group.
The infectiousness of the source $i$ at $t$ is given by,
\begin{equation*} 
f_B( b^{i}_t) = 
    \begin{cases}      
         \; 1, &\quad \text{ if } b^{i}_t \text{ is pre-severe or severe}   \\
         0.48, &\quad \text{ if } b^{i}_t \text{ is pre-mild or mild }   \\
         0.29, &\quad \text{ if } b^{i}_t \text{ is asymptomatic}   \\
         \; 0, &\quad \text{ otherwise}                 
    \end{cases}
\end{equation*}

and the strength of the interaction  between $i$ and $j$ at $t$ is defined as,

\begin{equation*} 
f_C(c^{ij}_t) = 
    \begin{cases}      
         2, &\quad \text{ if } c^{ij}_t\in H  \\
         1, &\quad \text{ otherwise.}                 
    \end{cases}
\end{equation*}

\begin{figure}[h!]
\centering
\includegraphics[width=0.4\textwidth]{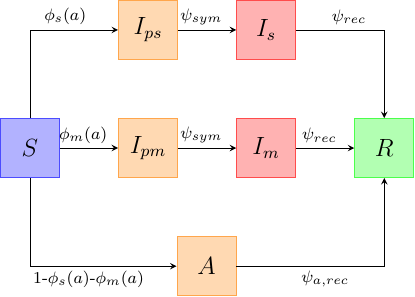}    
\caption{A diagram of the possible disease status evolution for an individual.}
\label{status_evolution}
\end{figure}

Upon infection,  an individual can be asymptomatic or develop mild or severe symptoms depending on her/his age group, see Figure \ref{status_evolution}. If the individual develops symptoms, she/he start by being in a pre-symptomatic state, in which she/he is infectious but have no symptoms. The $\phi_x(a)$ parameters are the probabilities of transition to a particular state depending on the age group of the individual. The $\psi_x$ parameters are gamma distributed and represent the time taken to make the transition.  Once an individual is recovered the model considers immunity. See Figure \ref{status_evolution} for a simplified schema of the possible transitions in the OpenABM-Covid19 model.

For more details on the OpenABM-Covid19 model see \href{https://github.com/aleingrosso/OpenABM-Covid19}{https://github.com/aleingrosso/OpenABM-Covid19} and \cite{hinch2021openabm}.

\section{Comparison with Mean-Field}\label{MF}

\begin{figure}[ht]
     \centering
     \begin{subfigure}[b]{0.2\textwidth}
         \centering
         \includegraphics[width=\textwidth]{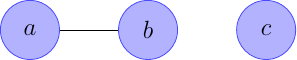}
         \caption{$t = 1$}
     \end{subfigure}
     \hfill
    \begin{subfigure}[b]{0.2\textwidth}
         \centering
         \includegraphics[width=\textwidth]{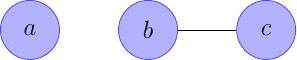}
         \caption{$t = 2$}
     \end{subfigure}
    \hfill
     \begin{subfigure}[b]{0.2\textwidth}
         \centering
         \includegraphics[width=\textwidth]{figures/Supplementary_material/time_1.pdf}
         \caption{$t = 3$}
     \end{subfigure}
     \hfill
     \begin{subfigure}[b]{0.2\textwidth}
         \centering
         \includegraphics[width=\textwidth]{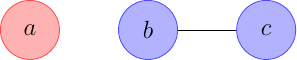}
         \caption{$t = 4$}
     \end{subfigure}
        \caption{Temporal interaction networks in a population of $N=3$. Individual $a$ is detected at day $t=4$.}
        \label{graph}
\end{figure}

We illustrate, using the toy example in Figure~\ref{graph}, the fundamental differences between our method ($2^\circ $CT) and the mean-field inference (MF) approach in \cite{baker2021epidemic}.  

First, we point out the difference in computing the risk of infection between the MF and $2^\circ $CT methods;  indeed, the risk in MF is updated for every individual day-by-day through an iterative process, while the risk  in $2^\circ $CT  is calculated at each day $t\geq t_0$ for only $1^\circ$contacts and $2^\circ$ contacts, by marginalising out over all the possible chains of transmission in the recent past. 

Second,  
we compare the two approaches through the example of  $3$ individuals $a$, $b$, and $c$, who interact during four days. The individual $a$ is detected at $t=4$ and is then considered as an index case in $[1:4]$.  We assume that $a$ is a patient zero, so the estimated time of infection is $\widehat{\tau}^a_{I}=0$.  The time of infection for individuals $b$ and $c$ are respectively $\widehat{\tau}^b_{I}=\infty$, and $\widehat{\tau}^c_{I}=\infty$, because they have not been detected yet.

\begin{figure}[ht]
     \centering
     \begin{subfigure}[b]{0.42\textwidth}
         \centering
         \includegraphics[width=\textwidth]{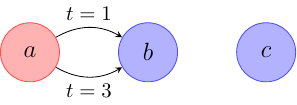}    
         \caption{Risk of $b$ at $t=4$}
     \end{subfigure}
     \hfill
    \begin{subfigure}[b]{0.42\textwidth}
         \centering
         \includegraphics[width=\textwidth]{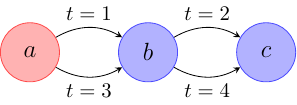}
         \caption{Risk of $c$ at $t=4$}
     \end{subfigure}
        \caption{Risky interactions taken into account in the $2^\circ $CT- method to compute: (a) the risk of $b$ at time $t=4$, (b) the risk of $c$ at time $t=4$}
\label{risk_propagation_2_CT}
\end{figure}

\begin{figure}[ht]
     \centering
     \begin{subfigure}[b]{0.42\textwidth}
         \centering
         \includegraphics[width=\textwidth]{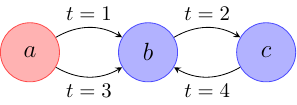}    
         \caption{Risk of $b$ at $t=4$}
     \end{subfigure}
     \hfill
    \begin{subfigure}[b]{0.42\textwidth}
         \centering
         \includegraphics[width=\textwidth]{figures/Supplementary_material/risk_c.pdf}    
         \caption{Risk of $c$ at $t=4$}
     \end{subfigure}
        \caption{Risky interactions taken into account in the MF method to compute: (a) the risk of $b$ at time $t=4$, (b) the risk of $c$ at time $t=4$.}
\label{risk_propagation_MF}
\end{figure}

Applying the two methods, we  determine  the risk of infection for $b$ and $c$ at $t=4$. In Figures \ref{risk_propagation_2_CT} and  \ref{risk_propagation_MF}, the arrows correspond to the risky interactions considered respectively by $2^\circ $CT and MF methods, to compute the risk of infection of $b$ and $c$ at time $t=4$. The values below/above the arrows correspond to the times at which these interactions take place.

We apply $2^\circ $CT method with  $\gamma =\zeta=4$. At time $t=4$, $b$ is a $1^\circ$contact of $a$, since they interacted at days $t=1,3$. As well, $c$ is $2^\circ$contact of $a$ since $c$ interacted with $b$ at days $t=2,4$. Hence, for the example presented in Fig. \ref{graph}, the $2^\circ $CT at time $t=4$ provides the following risk of infection of $b$ and $c$ 

\begin{displaymath}
\begin{split}
R^{b, 2^\circ}_{4,4}(4) & =   1 -  \Pro \left( \bigcap_{l=1}^4 Y^{ab}_l = 0 \; \middle| \;  \mathcal{O}^{4,4,2^\circ}_4 \right) \Pro \left( \bigcap_{l=1}^4 Y^{cb}_l = 0 \; \middle| \;  \mathcal{O}^{4,4,2^\circ}_4 \right)\\
   & =  1 - (1 - \lambda_{0, 1}^{a\rightarrow b})(1 - \lambda_{0, 3}^{a\rightarrow b}) 
\end{split}
\end{displaymath}

\begin{displaymath}
R^{c, 2^\circ}_{4,4}(4)  =   1 - \Pro \left( \bigcap_{l=1}^4 Y^{ac}_l = 0 \; \middle| \;  \mathcal{O}^{4,4,2^\circ}_4 \right) \Pro \left( \bigcap_{l=1}^4 Y^{bc}_l = 0 \; \middle| \;  \mathcal{O}^{4,4,2^\circ}_4 \right)
\end{displaymath}

where $\Pro \left( \bigcap_{l=1}^4 Y^{ac}_l = 0 \; \middle| \;  \mathcal{O}^{4,4,2^\circ}_4 \right) = 1$ and,  

\begin{displaymath}
\begin{split}
\Pro \left( \bigcap_{l=1}^4 Y^{bc}_l = 0 \; \middle| \;  \mathcal{O}^{4,4,2^\circ}_4 \right) & =  \sum_{d\in \{ 1, 2, 3, \infty\}} \Pro \left( \bigcap_{l=1}^4 Y^{bc}_l = 0 \; \middle| \;  \tau ^b_ I = d, \mathcal{O}^{4,4,2^\circ}_4 \right)\Pro \left(\tau ^b_ I = d  \; \middle| \;  \mathcal{O}^{4,4,2^\circ}_4 \right)\\
& =  \sum_{d\in \{ 1, 2, 3, \infty\}} g^b_{5,5}\left(d\right) \prod_{l=1}^4  \left( 1 - \lambda_{d, l}^{b\rightarrow c} \right)\\
&=  \lambda_{0, 1}^{a\rightarrow b} \prod_{l\in \{2,4\}}  ( 1 - \lambda_{1, l}^{b\rightarrow c} ) + \lambda_{0, 3}^{a\rightarrow b}(1 - \lambda_{0, 1}^{a\rightarrow b})(1-\lambda_{3, 4}^{b\rightarrow c})+ \prod_{l\in \{1,3\}} (1 - \lambda_{0, l}^{a\rightarrow b}). \\
\end{split}
\end{displaymath}

For the MF method, we suppose that the  time between infection and testing is greater than $4$, and we set $t_{MF}=4$, ($t_{MF}$ stands for the integration time of the MF process). Once $a$ is detected, to compute the MF risk we assume that the initial risk of $a$ is 1, while for $b$ and $c$ it is considered as 0.  Therefore, the MF risk of the individual $b$ at time $t= 4 $ is

\begin{displaymath}
\begin{split}
R_{\text{MF}}^b(4) 
& = 1 - (1 - p_{ 1}^{a\rightarrow b})(1 - p_{ 3}^{a\rightarrow b})(1 - p_{ 4}^{c\rightarrow b}p_{2}^{b\rightarrow c}p_{ 1}^{a\rightarrow b}) \\
& =  1 - \Bigl(1 - R_{\text{MF}}^b(3) \Bigr)\Bigl(1 - p^{c\rightarrow b}_4  R_{\text{MF}}^c(3) \Bigr), \\
\end{split}
\end{displaymath}

where $p_t^{i\rightarrow j}$ is the probability that $i$ infects $j$ at time $t$, which does not depend on the time of infection of the source $i$. 

The MF risk of the individual $c$ at time $t=4 $ is

\begin{displaymath}
\begin{split}
R_{\text{MF}}^c(4) 
& = 1 - \left( 1 - p_{ 1}^{a\rightarrow b}p_{ 2}^{b\rightarrow c}\right)
    \left[1 - p_{ 4}^{b\rightarrow c}\left(1 - (1 - p_{ 1}^{a\rightarrow b})(1 - p_{ 3}^{a\rightarrow b})\right)\right] \\
& =  1 - \Bigl(1 - R_{\text{MF}}^c(3) \Bigr)\Bigl(1 - p^{b\rightarrow c}_4  R_{\text{MF}}^b(3) \Bigr). \\
\end{split}
\end{displaymath}

Notice that in the MF method, the risk is propagated ``bidirectionally'', meaning that it is not only ``forward'' in the direction of the transmission  given the observations, as the proposed $2^\circ$CT method. Hence, whenever two individuals interact, they interchange and upgrade their risk regardless where their infection came from, and assuming independence in all possible transmission chains. For example, in Figure~\ref{graph} at time $t=2, 4$, individuals $b$ and $c$ are in contact and interchange their risk. As a consequence, the risk of $b$ is increased by the current risk of $c$, but the risk carried by $c$ came from individual $b$. This is what we call a \textit{cycling back} effect, which can cause an artificial increase of the risk for some individuals in the population. 
 Moreover, in some cases it allows the risk of $c$ to be greater than the risk of $b$, that is $R_{\text{MF}}^{b}(4) \leq R_{\text{MF}}^{c}(4)$.
 In the $2^\circ$CT method, we avoid this effect and guarantee that, for any possible transmission probability function per interaction, we have that $R^{b, 2^\circ}_{4,4}(4) \geq   R^{c, 2^\circ}_{4,4}(4) $.

\end{document}